\newif\ifconference
\conferencefalse

\newif\ifjournal
\journaltrue

\documentclass[UKenglish, a4paper, runningheads, envcountsect, envcountsame]{llncs}

\usepackage{jbw-debug}
\usepackage{hyperref}
\usepackage{jbw-fix-hyperref-autoref}
\usepackage{autofe}
\makeatletter
  \input{t1enc.def}%
  \input{ot1enc.def}%
\makeatother

\ifconference
  
\else
  
\fi

\usepackage{jbw-thm}
\usepackage{jbw-boxfigure}
\usepackage{todonotes}
\usepackage{outlines}
\usepackage{array}
\usepackage{xcolor}
\usepackage{amsmath}
\usepackage{amsfonts}
\usepackage{amssymb}
\usepackage{datetime}%
\usepackage{upgreek}
\usepackage{colonequals}
\usepackage{environ}
\usepackage{centernot}
\usepackage{proof}
\usepackage{stmaryrd}
\usepackage[f]{esvect}

\usepackage{scalerel}
\usepackage{accsupp}

\newcommand*{\llbrace}{%
  \BeginAccSupp{method=hex,unicode,ActualText=2983}%
    \textnormal{\usefont{OMS}{lmr}{m}{n}\char102}%
    \mathchoice{\mkern-4.05mu}{\mkern-4.05mu}{\mkern-4.3mu}{\mkern-4.8mu}%
    \textnormal{\usefont{OMS}{lmr}{m}{n}\char106}%
  \EndAccSupp{}%
}

\newcommand*{\rrbrace}{%
  \BeginAccSupp{method=hex,unicode,ActualText=2984}%
    \textnormal{\usefont{OMS}{lmr}{m}{n}\char106}%
    \mathchoice{\mkern-4.05mu}{\mkern-4.05mu}{\mkern-4.3mu}{\mkern-4.8mu}%
    \textnormal{\usefont{OMS}{lmr}{m}{n}\char103}%
  \EndAccSupp{}%
}

\newcommand{\mset}[1]{\llbrace\,#1\,\rrbrace}

\newcommand{\msf}{\mathsf}
\newcommand{\labelcolour}{darkgray!70!blue}
\newcommand{\mathlabel}[1]{{\color{\labelcolour}{(\textsc{#1})}}}

\DeclareRobustCommand{\riota}{\rotatebox[origin=C]{180}{$\iota$}}
\newcommand{\binddot}[3]{#1\:#2.\,#3}
\newcommand{\ccolon}{\mathrel{\overline{\mbox{::\rule{0pt}{1.5mm}}}}}

\newcommand{\lam}[2]{\binddot{\lambda}{#1}{#2}}
\newcommand{\app}[2]{#1\,#2}
\newcommand{\apptwo}[3]{#1\,#2\,#3}
\newcommand{\appthree}[4]{#1\,#2\,#3\,#4}
\newcommand{\appfour}[5]{#1\,#2\,#3\,#4\,#5}
\newcommand{\appldots}[3]{#1\,#2\ldots#3}
\newcommand{\domain}[1]{\mathsf{d}_{#1}}

\newcommand{\TVar}{\mathsf{TVar}}
\newcommand{\Type}{\mathsf{Type}}

\newcommand{\TV}{\mathsf{TV}}
\newcommand{\typ}{\msf{typ}}
\newcommand{\RVar}{\mathsf{RVar}}
\newcommand{\Var}{\mathsf{Var}}

\newcommand{\Const}{\mathsf{Const}}
\newcommand{\instOf}{\mathrel{\lessdot}}

\newcommand{\bool}{\star}
\newcommand{\Fun}{\mathbin{\Rightarrow}}
\newcommand{\tvar}[1]{\mathsf{tv}_#1}
\newcommand{\var}[1]{\mathsf{v}_#1}

\newcommand{\ctyp}{\mathsf{ctyp}}

\newcommand{\typSub}[1]{\llbracket#1\rrbracket}
\newcommand{\Term}{\mathsf{Term}}
\newcommand{\RTerm}{\Term_{\mathsf{R}}}
\newcommand{\ATerm}{\Term_{\upalpha}}
\newcommand{\FV}{\mathsf{FV}}
\newcommand{\st}{\msf{set}}
\newcommand{\trm}{\msf{trm}}

\newcommand{\ZFplus}{\mathsf{ZF}^+}
\newcommand{\elem}{\mathrel{\dot{\epsilon}}}
\newcommand{\cceq}{\coloncolonequals}

\newcommand{\abbreviates}{\mathrel{\colonequals}}
\newcommand{\metain}{\mathrel{\epsilon}}
\newcommand{\notmetain}{\centernot\metain}
\newcommand{\meq}{\mathrel{\equiv}}
\newcommand{\nat}{\mathbb{N}}

\newcommand{\nil}{{\diamond}}
\newcommand{\mlist}[1]{[\,#1\,]}
\newcommand{\listin}{\metain:}

\newcommand{\cons}{\mathbin{\#}}
\newcommand{\append}{\mathbin{@}}

\newcommand{\proves}{\vdash}
\renewcommand{\lnot}[1]{\app\neg{#1}}
\newcommand{\imp}{\rightarrow}
\renewcommand{\iff}{\leftrightarrow}
\newcommand{\ifconst}{\mathsf{IF}}

\newcommand{\all}[2]{\binddot{\forall}{#1}{#2}}
\newcommand{\ex}[2]{\binddot{\exists}{#1}{#2}}
\newcommand{\exOneSym}{\exists!}
\newcommand{\exOne}[2]{\binddot{\exOneSym}{#1}{#2}}
\newcommand{\atMostOneSym}{\exists_{\leq 1}}
\newcommand{\atMostOne}[2]{\binddot{\atMostOneSym}{#1}{#2}}
\newcommand{\defdes}[3]{
  \binddot{\riota}{#1}{#2\,\mathsf{else}\,#3}}
\newcommand{\ifelse}[3]{
  \mathsf{if}\:#1\:\mathsf{then}\:#2\:\mathsf{else}\:#3}

\newcommand{\SoftTypes}{\mathsf{SoftTypeOps}}

\newcommand{\fun}{\rightarrowtriangle}
\newcommand{\inter}{\mathbin\sqcap}
\newcommand{\uni}{\mathbin{\sqcup}}
\newcommand{\subtyp}{\sqsubseteq}
\newcommand{\depfun}[2]{\binddot{\Uppi}{#1}{#2}}
\newcommand{\lamelse}[3]{\lam{#1}{#2\:\mathsf{else}\:#3}}

\newcommand{\cla}{\mathcal{C}}

\newcommand{\cladeps}{\mathcal{D}}
\newcommand{\claconsts}{\mathcal{K}}

\newcommand{\params}{\mathsf{params}}
\newcommand{\axioms}{\mathsf{axioms}}
\newcommand{\defs}{\mathsf{defs}}
\newcommand{\tv}{\mathsf{tv}}

\newcommand{\consts}{\mathsf{consts}}
\newcommand{\class}{\mathsf{class}}

\newcommand{\logo}{\mathsf{logo}}
\newcommand{\cargo}{\mathsf{cargo}}
\newcommand{\feat}{\mathcal{F}}
\newcommand{\featG}{\mathcal{G}}
\newcommand{\defaultval}[1]{\circ_{#1}}

\newcommand{\GZF}{\mathsf{GZF}}
\newcommand{\Set}{\mathsf{Set}}
\newcommand{\SetMem}{\mathsf{SetMem}}

\newcommand{\cSetOf}{\mathsf{SetOf}}
\newcommand{\SetOf}[1]{\app\cSetOf{#1}}

\newcommand{\cUnion}{\bigcup}
\newcommand{\Union}[1]{\app\cUnion{#1}}

\newcommand{\cPow}{\mathcal{P}}
\newcommand{\Pow}[1]{\app\cPow{#1}}

\newcommand{\cRepl}{\mathsf{Repl}}
\newcommand{\Repl}[2]{\apptwo\cRepl{#1}{#2}}

\newcommand{\Emp}{\emptyset}

\newcommand{\cSucc}{\mathsf{Succ}}
\newcommand{\Succ}[1]{\app\cSucc{#1}}

\newcommand{\Inf}{\mathsf{Inf}}
\newcommand{\cReplPred}{\mathsf{ReplPred}}
\newcommand{\ReplPred}[1]{\app\cReplPred{#1}}

\newcommand{\setof}[1]{\{\,#1\,\}}
\newcommand{\upair}{\mathsf{upair}}

\newcommand{\Ord}{\mathsf{Ord}}
\newcommand{\csucc}{\mathsf{succ}}
\renewcommand{\succ}[1]{\app\csucc{#1}}
\newcommand{\Limit}{\mathsf{Limit}}

\newcommand{\cOrdRec}{\mathsf{OrdRec}}
\newcommand{\OrdRec}[4]{\appfour\cOrdRec{#1}{#2}{#3}{#4}}
\newcommand{\cpredSet}{\mathsf{predSet}}
\newcommand{\predSet}[1]{\app\cpredSet{#1}}

\newcommand{\Func}{\mathsf{Fun}}
\newcommand{\capp}{\mathsf{app}}
\newcommand{\fapp}[3]{\appthree\capp{#1}{#2}{#3}}

\newcommand{\cmkfun}{\mathsf{mkFun}}
\newcommand{\mkfun}[2]{\apptwo\cmkfun{#1}{#2}}

\newcommand{\cdom}{\mathsf{dom}}
\newcommand{\dom}[1]{\app\cdom{#1}}

\newcommand{\cran}{\mathsf{ran}}
\newcommand{\ran}[1]{\app\cran{#1}}

\newcommand{\FunMem}{\mathsf{FunMem}}
\newcommand{\cFunPred}{\mathsf{FunPred}}
\newcommand{\FunPred}[1]{\app\cFunPred{#1}}
\newcommand\pfun{\mathbin{\ooalign{\hfil$\mapstochar\mkern5mu$\hfil\cr$\to$\cr}}}

\newcommand{\prettypair}[2]{\langle{#1},{#2}\rangle}
\newcommand{\Pair}{\mathsf{Pair}}

\newcommand{\cpair}{\mathsf{pair}}
\newcommand{\pair}[2]{\apptwo\cpair{#1}{#2}}

\newcommand{\PairMem}{\mathsf{PairMem}}

\newcommand{\csupOrd}{\mathsf{supOrd}}
\newcommand{\supOrd}[1]{\app\csupOrd{#1}}

\newcommand{\exc}{\bullet}

\newcommand{\Ordinal}{\mathsf{Ordinal}}
\newcommand{\OrdinalRec}{\mathsf{OrdinalRec}}
\newcommand{\OPair}{\mathsf{OPair}}
\newcommand{\Function}{\mathsf{Function}}
\newcommand{\Exc}{\mathsf{Exc}}

\newcommand{\bfGZF}{\textbf{GZF}}
\newcommand{\bfOrdinal}{\textbf{Ordinal}}
\newcommand{\bfOrdRec}{\textbf{OrdRec}}
\newcommand{\bfFunction}{\textbf{Function}}
\newcommand{\bfOPair}{\textbf{OPair}}
\newcommand{\Exception}{\mathsf{Exception}}
\newcommand{\bfExc}{\textbf{Exc}}

\newcommand{\default}{\mathsf{default}}
\newcommand{\typList}{\mathsf{typList}}
\newcommand{\otherwise}{\mathsf{otherwise}}
\newcommand{\allOtherwise}{\mathsf{allOtherwise}}

\newcommand{\cover}{\mathsf{cover}}
\newcommand{\disjoint}{\mathsf{disjoint}}

\newcommand{\admitCargo}{\mathsf{admitCargo}}
\newcommand{\restrictCargo}{\mathsf{restrictCargo}}
\newcommand{\cargoAx}{\mathsf{cargoAx}}
\newcommand{\blist}{\mathcal{B}}
\newcommand{\wlist}{\mathcal{W}}

\newcommand{\GST}{\mathsf{GST}}
\newcommand{\mkGST}{\mathsf{mkGST}}

\newcommand{\Tagging}{\mathsf{Tagging}}
\newcommand{\ModelBase}{\mathsf{ModelBase}}

\newcommand{\cTier}{\mathsf{Tier}}
\newcommand{\Tier}[1]{\app\cTier{#1}}
\newcommand{\cTierImp}{\mathsf{TierImp}}
\newcommand{\TierImp}[3]{\appthree{\cTierImp\,}{#1\,}{#2\,}{#3}}
\newcommand{\Excluded}{\mathsf{Excluded}}
\newcommand{\Tag}{\mathsf{Tag}}
\newcommand{\tagmap}{\mathbin\oplus}
\newcommand{\cdisj}{\biguplus}
\newcommand{\disj}[1]{\app\cdisj{#1}}
\newcommand{\M}{\mathbb{M}}
\newcommand{\Ex}{\mathsf{E}}
\newcommand{\zero}{\mathsf{zero}}
\newcommand{\limit}{\mathsf{limit}}

\newcommand{\constructor}{\mathfrak{C}}
\newcommand{\mcompvar}{\mathcal{M}}
\newcommand{\ttag}{\mathsf{tag}}
\newcommand{\mcomp}{\mathsf{mcomp}}
\newcommand{\cfst}{\mathsf{fst}}
\newcommand{\fst}[1]{\app\cfst{#1}}
\newcommand{\csnd}{\mathsf{snd}}
\newcommand{\snd}[1]{\app\csnd{#1}}

\newcommand{\mGZF}{\mathsf{mGZF}}
\newcommand{\set}{\mathsf{set}}
\newcommand{\kons}{\mathsf{struc}}

\newcommand{\mSet}{\mathsf{mSet}}
\newcommand{\mSetMem}{\mathsf{mSetMem}}

\newcommand{\mmem}{\mathrel{\overline\in}}

\newcommand{\cmSetOf}{\mathsf{mSetOf}}
\newcommand{\mSetOf}[1]{\app\cmSetOf{#1}}

\newcommand{\cmUnion}{\overline\bigcup}

\newcommand{\cmPow}{\overline{\mathcal{P}}}
\newcommand{\mPow}[1]{\app\cmPow{#1}}

\newcommand{\cmReplPred}{\mathsf{mReplPred}}
\newcommand{\mReplPred}[1]{\app\cmReplPred{#1}}
\newcommand{\cmRepl}{\overline{\mathcal{R}}}
\newcommand{\mRepl}[2]{\apptwo\cmRepl{#1}{#2}}

\newcommand{\mEmp}{\overline\emptyset}

\newcommand{\cmSucc}{\mathsf{mSucc}}
\newcommand{\mSucc}[1]{\app\cmSucc{#1}}

\newcommand{\mSetOrd}{\mathsf{mSetOrd}}
\newcommand{\mInf}{\mathsf{mInf}}
\newcommand{\excluded}{\mathsf{ex}}
\newcommand{\map}{\mathsf{map}}
\newcommand{\translate}{\mathsf{translateAxioms}}
\newcommand{\resp}{\mathsf{respThms}}

\newcommand{\leftsub}{\mathbin{-_{\msf{left}}}}
\newcommand{\lamfun}[2]{\binddot{\dot\lambda}{#1}{#2}}

\iftrue
  \makeatletter
    \renewcommand\section{\@startsection{section}{1}{\z@}%
                           {-15\p@ \@plus -4\p@ \@minus -4\p@}%
                           {10\p@ \@plus 4\p@ \@minus 4\p@}%
                           {\normalfont\large\bfseries\boldmath
                            \rightskip=\z@ \@plus 8em\pretolerance=10000 }}
    \renewcommand\subsection[1]{\@startsection{subsection}{2}{\z@}%
                         {-12\p@ \@plus -4\p@ \@minus -4\p@}%
                         {-0.5em \@plus -0.22em \@minus -0.1em}%
                         {\normalfont\normalsize\bfseries\boldmath}{#1.}}
    \renewcommand\normalsize{%
       \@setfontsize\normalsize\@xpt\@xiipt
       \abovedisplayskip 6\p@ \@plus2\p@ \@minus3\p@
       \abovedisplayshortskip \z@ \@plus3\p@
       \belowdisplayshortskip 6\p@ \@plus2\p@ \@minus3\p@
       \belowdisplayskip \abovedisplayskip
       \let\@listi\@listI}
    \normalsize
  \makeatother
\fi

\begin{document}

\title{Isabelle/HOL/GST: A Formal Proof Environment for Generalized Set Theories}
\titlerunning{Isabelle/HOL/GST}
\author{Ciar\'an Dunne \and J. B. Wells}
\institute{Heriot-Watt University}
\maketitle

\begin{abstract}
  \setlength{\parskip}{1ex}%
  A \emph{generalized set theory} (GST) is like a standard set theory but also can have non-set
  structured objects that can contain other structured objects including sets.
  This paper presents Isabelle/HOL support for GSTs, which are treated as type classes that combine
  \emph{features} that specify kinds of mathematical objects, e.g., sets, ordinal numbers,
  functions, etc.
  GSTs can have an exception feature that eases representing partial functions and undefinedness.
  When assembling a GST, extra axioms are generated following a user-modifiable policy to fill
  specification gaps.
  Specialized type-like predicates called \emph{soft types} are used extensively.
  Although a GST can be used without a model, for confidence in its consistency we build a model
  for each GST from components that specify each feature's contribution to each tier of a
  von-Neumann-style cumulative hierarchy defined via ordinal recursion, and we then connect the
  model to a separate type which the GST occupies.
\end{abstract}

\section{Introduction}

\subsection{Set Theory}

Many mathematicians (but not all) have long regarded set theories as
suitable foundations of mathematics, in particular Zermelo/Fraenkel set theory (ZF) and other
related theories, e.g., ZF plus the Axiom of Choice (ZFC) and
Tarski/Grothendieck (TG) set theory which adds a universe axiom.
At its core, ZF is a domain type $V$ whose members are called ``sets'', a predicate $\in$
(membership) of type
$V \Fun V \Fun \msf{bool}$, and axioms specifying $\in$.
Due to cardinality constraints, some operators like $\in$ can not themselves
be members of the domain $V$ but must live in higher types.
Although just the predicate $\in$ is enough, formalizing ZF is easier with some constants and
additional operators at a few further types, e.g., $\cPow$ (power set) and $\cUnion$ (union) at type $V \Fun V$ and
$\cRepl$ (replacement) at type $V \Fun (V \Fun V \Fun \msf{bool}) \Fun V$.
Nonetheless, ZF needs few types and nearly all interesting mathematical objects live in type $V$.

ZF is usually specified by ``axioms'' written in first-order logic (FOL), but ZF can also be given
in higher-order logic (HOL).
Actively used proof systems implementing ZF or TG in FOL include Isabelle/ZF, Mizar (TG), and
Metamath (ZF, when using the set.mm database).
Active systems in HOL include the Isabelle/HOL development ``ZFC in
HOL''~\cite{Pau:ZFC_in_HOL-AFP:2019}.
Other systems include:
Isabelle/Set~\cite{Kap+Che+Kra:Isabelle-Set-git-2022-05} (TG, HOL),
Isabelle/Mizar \cite{Kal+Pak:JAR-2019} (TG),
and
Egal~\cite{Bro+Kal+Pak:ITP-2019} (TG, HOL).

In all these systems, except for a few key operators like $\in$, every mathematical object is a set.
Arrangements of sets represent numbers, ordered pairs, functions, and nearly all other kinds of
mathematical objects.
Because sets are used to represent everything, representation overlaps are unavoidable.
With the most commonly used representations, sad coincidences include, e.g., that
$\prettypair 01=\{1,2\}$ where $0$, $1$, and $2$ are natural numbers, and that the successor
function on natural numbers is equal to the number $1$ as an integer.
This troubles philosophers, leads university teachers to choose to deceive students, makes correct
definitions more challenging, and adds difficulty to formalization.

\subsection{Higher-Order Logic}

Although many mathematicians favor set theory as a foundation of mathematics, many computer system
implementers prefer formalisms where numerous types are used, rather than set theory's ``one big
type''.
Most of these systems extend Church's $\lambda$-calculus with simple types.
We consider here HOL, which needs at its core only one type constructor $\Fun$ and axioms and
inference rules for constants for equality ($=$) and implication ($\imp$).
Most connectives ($\land$, $\neg$, $\all$, etc.)\ are added via simple definitions that provide
convenience, compactness, and readability, but no extra power.
Sometimes domain-specific axioms are added that do provide extra power (which raises the question of
whether these extensions are consistent, which we will address for our systems later in this paper).
The systems we present in this paper are developed in Isabelle/HOL, the most active and widely used
HOL proof system.
Isabelle/HOL adds locale and type class mechanisms that support modularity, some limited type
polymorphism, and overloading, and adds type definition features such as semantic subtypes, quotient
types, and (co)recursive datatypes.
Although Isabelle/HOL can be used just as a logical framework in which to define a set theory (as
this paper does),
typical uses of Isabelle/HOL generally put different kinds of mathematical objects into numerous
fine-grained types.

\subsection{Types}

Consider ways ``types'' can be useful.
One view of ``type'' is as an aspect of some object that can be inspected to help determine what you
\emph{should or want to} do with it.
For example, when storing an incoming parcel, you might store food somewhere cold and jewelry in a
safe.
This corresponds to using types for overloading.
Another view of ``type'' is as determining what you \emph{meaningfully can} do with some object.
For example, if you have two pieces of paper you want to attach, ice cream that you want to eat, a
stapler, some staples, a bowl, and a spoon, you will want to use the right tools for each task.
This corresponds to using types for avoiding meaningless combinations.
There is no firm boundary between these two notions, e.g., you might \emph{want} to store food
somewhere cold because otherwise it might rot making the storage \emph{not meaningful}.

Formal proving needs both views of ``types'' and these views interact.
Using a lemma usually requires knowing if an operation is ``defined'' and the result's ``type'', and
the answers to the same questions for operations on results \emph{ad infinitum}.
It is also frequently very useful to make decisions based on ``types''.

With this in mind, compare the set theory and HOL approaches (excluding HOL used solely
as a framework for a set theory).
HOL typically has precise and (usually) useful types for each object,
obtained automatically with (often) reasonable efficiency, but sometimes it is quite hard to find
types that allow fitting everything that is needed together.
In set theory, typically the type of nearly everything is $\msf{set}$, which
conveys little information, but sets are a rich collection of every imaginable predicate and are
usable as ``types''.
However, there is (with currently available systems) a lack of automation for
finding the right ``types'' and making them available in the right places at the right
times.

\subsection{Generalized Set Theories}

A \emph{generalized set theory} (GST)~\cite{aczel} specifies a domain type that contains set
objects and may also contain other kinds of non-set structured objects, e.g., non-set
ordered-pairs, non-set functions, non-set relations, etc.
Many mathematicians view their work as based on ZF but their practice is often inconsistent with
numbers, tuples, functions, etc., actually being sets.
We believe much mathematics is, in effect, actually using a GST.

Subtype distinctions within a GST's domain type avoid representation overlaps.
In a GST with non-overlapping sets, numbers, and ordered pairs, the answer to ``is the ordered pair
$\prettypair 01$ equal to the set $\{1,2\}$?'' is ``no'', an improvement over ``yes'' or ``maybe''
or ``how dare you ask that question!''.
Subtype distinctions also help with showing that operations have meaningful results, and with
choosing the most appropriate reasoning for each object.

A GST can have an exception object $\bullet$ (spoken ``boom'') which can not be
confused with any other kind of object, can not lurk hidden inside any other object, and is useful
for representing ``undefined'' results.
This is similar to how terms in free logic can be ``undefined'' or fail to denote, but avoids
the costs of actual undefinedness, e.g., inability to use standard HOL proof systems.
This also avoids problems of other undefinedness approaches, e.g., the weakness of three-valued
logic, the confusion caused by defining $1/0$ to be $0$, the need for amazingly complicated function
domain specifications to handle definedness gaps, etc.
Our approach avoids needing separate inner/outer domain quantifiers.

Recent unformalized theoretical work by us with Kamareddine~\cite{Dun+Wel+Kam:CICM-2021} showed how
to combine \emph{features} to make a GST.
Features include what are traditionally called \emph{structures}, e.g., the natural numbers, and
also include “large” (proper-class-sized) concepts like the sets, the ordered pairs, the functions,
and the ordinals.
This approach, also followed in this paper, is as follows.
Each GST has a single domain type $\domain{}$ and some features.
Each feature specifies the existence of some objects in $\domain{}$, which the feature is usually
considered to ``own'', and can specify relationships among all the objects in $\domain{}$.
Additional axioms
specify that every object is ``owned'' by exactly one feature.
Features are specified as independently as is feasible so, e.g., the specification of ordered pairs
knows nothing about sets and \emph{vice versa}.
This is useful because, e.g., for sets and ordered pairs,
(1) mathematicians think of these as distinct kinds of things,
(2) the independent specification of ordered pairs is easier to comprehend
than, e.g., Kuratowski's definition in terms of sets, and
(3) this enables custom GSTs with only the desired primitive features.
Our recent work also suggested theoretically how to build a model of a GST $Q$ within another GST
$P$ and then connect that model to a type containing the GST $Q$, thereby reducing the question of
the consistency of $Q$ to the consistency of $P$.
Our previous work had not, until this paper, been formalized.%
\footnote{A mostly formalized proof had been given earlier in Isabelle/ZF of the existence of a model
  of a much simpler system with just two fixed features~\cite{dunne2020}.}
Aside from this paper and the work mentioned above, the only model building for GSTs seems to be by
Aczel and Lunnon~\cite{aczel1991universes}, and their set theories have the anti-foundation axiom,
which is quite different from our work.

\subsection{Isabelle/HOL/GST}

Building on Isabelle/HOL and earlier work on GSTs,
we present a formal proving environment
for defining GSTs and reasoning about them, including building models for them.
The development consists of much Isabelle/Isar code for high-level formalization,
as well as Isabelle/ML for the implementation of machinery that assists
GST specification and model building.
The source code of our development and instructions for use
 are available at \url{https://www.macs.hw.ac.uk/~cmd1/isabelle-gst/}.

Our development provides:
(1)~Definitions of a number of \emph{soft type constructors}, operators that
construct and manipulate \emph{soft types}, HOL predicates of type $\tau\Fun\bool$, where $\bool$ is
this paper's name for $\msf{bool}$, and proofs of many useful properties of these.
(2)~A formal notion of \emph{feature}, with a corresponding implementation as an Isabelle/ML record type.
A feature contains a pointer to an Isabelle (type) class,
which manages an abstract specification of the
dependencies on other classes, signature, axioms, definitions, theorems, syntax, etc., of a feature,
and also contains information used when combining features.
(3)~Definitions and theorems in Isabelle theories for features for ZF-style sets, functions,
ordinals, ordinal recursion, an exception object, ordered pairs, natural numbers and binary
relations.%
\footnote{%
  Natural numbers and binary relations can be found in the Isabelle source code, as can
  ordered pairs, which also have a \LaTeX\ presentation in this paper's long version.}
(4)~Automation for combining features to define GSTs,
which are implemented as Isabelle classes with axioms for enforcing:
(a)~that each object in the domain of individuals is owned by exactly one feature,
(b)~a policy specifying allowed combinations of objects of different features, and a
policy for filling in specification gaps.
These two policies primarily support approaches to handling
``undefined'' operations, including the use of an exception object.
(5)~Generic definitions and reasoning for models of GSTs,
which are cumulative hierarchies defined using ordinal recursion.
(6)~A notion of \emph{model component}, with a corresponding implementation as an ML record type.
Each model component specifies a schematic description of a part of a model,
primarily constraints placed
on the zero, successor, and limit cases of the ordinal recursion
used in defining models.
(7)~Automatic generation of terms, theorems and proof states that assist
     in implementing GST models.
(8)~Automatic lifting and generation of transfer rules for constants on types obtained by
Isabelle/HOL type definitions on GST models, gained by interfacing Lifting and Transfer.
(9)~A bootstrap of our development from Paulson's ``ZFC in HOL'',
carried out by instantiating our $\GZF$ (set), $\Ordinal$, $\OrdinalRec$ (ordinal recursion),
$\OPair$ (ordered pair), and $\Function$ classes at the type $\mathbf{V}$ of ``ZFC in HOL'', which
makes a GST in which every
object is a set.
(10)~An example GST called $\ZFplus$ (defined as a class)
   with sets and all of the following as non-set objects:
     functions, ordinals, and the exception object.
(11)~An Isabelle type $\domain 0$ which instantiates the $\ZFplus$ class, obtained by type
definition on a model built using our model-building kit in $\mathbf{V}$, justifying confidence in
using $\ZFplus$, provided you have faith in Isabelle/HOL and the axioms of ZFC that were added at
$\mathbf{V}$.

Items (5), (6), (7), and (8) with model building details are not presented in this paper, but are
described in the appendix of the long version of this paper.
Full details can be found in the \texttt{ModelKit/} directory of the Isabelle source.

\subsection{Summary of Contributions}
\label{sec:summ-contr}

This paper presents a formalization of generalized set theories in Isabelle/HOL.
\Autoref{sec:math-defin-logic} formulates a logical framework with classes like those of Isabelle.
\Autoref{sec:gst-features} defines GST features as classes with associated data.
\Autoref{sub:example-features} presents example features for ZF-style sets, ordinals,
functions, and an exception element.
\Autoref{sec:combination} defines how to combine features to create a GST and shows how to combine
our example features to make the GST $\ZFplus$ which has all these features within one
domain type.
\Autoref{sec:examples-working-gst} presents examples of working in $\ZFplus$.
\Autoref{sec:zfc-zfplus} presents a methodology for building models of GSTs in Isabelle and how we
use the Lifting and Transfer packages to create types that instantiate GSTs.
\Autoref{sec:zfc-zfplus} also discusses how we justify confidence in
$\ZFplus$ by building a model in the type $\mathbf{V}$ of ``ZFC in HOL'', defining a type $\domain0$
isomorphic to this model, and instantiating $\ZFplus$ at $\domain 0$.

\newcommand{\BaseSyntax}{
  \begin{array}{@{}r@{\ \elem\ }l@{}l@{\qquad}r@{\ \elem\ }l@{}l@{}}
        i,j,k,n,m
      & \nat
      &
    &
        \bar x
      & \RVar
      & \,\cceq\,
          \var 1 \mid \var 2 \mid \var 3 \mid \cdots
  \\[.5ex]
        \alpha, \beta
      & \TVar
      & \,\cceq\,
          \tvar 1 \mid \tvar 2 \mid \tvar 3 \mid \cdots
    &
        \sigma, \tau, \rho
      & \Type
      & \,\cceq\,
          \alpha \mid \bool \mid \domain i
          \mid \sigma \Fun \tau
  \\[.5ex]
        \bar\kappa, \bar\ell
      & \Const
      & \multicolumn{4}{l}{
        \,\cceq\,
               {\imp} \mid {=} \mid \True \mid \False
          \mid \neg \mid {\land} \mid {\lor}
          \mid \forall \mid \exists \mid \exOneSym
          \mid \atMostOneSym \mid \riota \mid \ifconst
          \mid \cdots
        }
  \\[.5ex]
    \multicolumn{6}{l}{
        b,c,d,p,q,u,v,x,y,z
      \ \elem\
        \Var
      \ \cceq\
        [\bar x,\sigma]
    }
  \end{array}}

\newcommand{\TermSyntax}{
  \bar B, \bar C
  \elem \RTerm \cceq
    x \mid \kappa
      \mid \app{\bar B}{\bar C}
      \mid \lam{x}{\bar B}
  \qquad\qquad
  B, C, F, G, P, Q \elem \Term
}

\newcommand{\vTypeRule}{\infer{{\bar x}_\sigma :: \sigma}{}}

\newcommand{\cTypeRule}{\infer{\bar\kappa_\sigma :: \sigma}
                              {\sigma \instOf \ctyp(\bar\kappa)}}

\newcommand{\lamTypeRule}{
  \infer{(\lam{x}{B}) :: \sigma \Fun \tau}
        {x :: \sigma; & B :: \tau}}

\newcommand{\appTypeRule}{
  \infer{(\app B C) :: \tau}
        {B :: \sigma \Fun \tau; & C :: \sigma}}

\newcommand{\FreeTypeVariables}{
  \begin{array}{l@{\,\meq\,}l@{\quad}l@{\,\meq\,}l@{\quad}l@{\,\meq\,}l}
      \TV_\typ(\alpha) & \mset{\alpha}
    &
      \TV_\typ(\bool) \meq \TV_\typ(\domain i) & \varnothing

    &
      \TV_\typ(\sigma\Fun\tau) & \TV_\typ(\sigma) \Cup \TV_\typ(\tau)
  \end{array}}

\newcommand{\FreeVariables}{
  \begin{array}{l@{\,\meq\:}l}
      \FV(x) & x
  \\
      \FV(\app{\hat B}{\hat C}) & \FV(\hat B) \Cup \FV(\hat C)
  \\
      \FV(\lam{x}{\hat B}) & \FV(\hat B) \bbslash \mset{x}
  \end{array}}

\newcommand{\InferenceRules}{%
  \begin{array}{c@{\,\,}l}
    \mathlabel{assm} &
    \text{If $\varphi \metain \HOL \Cup \ZFCinHOL \Cup \Updelta \Cup \Gamma$, 
          then $\Updelta; \Gamma \proves \varphi$}
  \\
    \mathlabel{impI} &
    \text{If $\Updelta; \Gamma \cup \{\varphi\} \proves \psi$,
          then $\Updelta; \Gamma \proves \varphi \imp \psi$}
  \\
    \mathlabel{impE}
  &
    \text{If $\Updelta; \Gamma \proves \varphi \imp \psi$
          and $\Updelta; \Gamma \proves \varphi$,
          then $\Updelta; \Gamma \proves \psi$}
  \\
    \mathlabel{typ-inst}
  &
    \text{If $\Updelta; \Gamma \proves \varphi$ and
             $\alpha \notmetain \TV_\set(\Gamma)$,
          then $\Updelta; \Gamma \proves \varphi[\alpha := \sigma]$}
  \\
    \mathlabel{trm-inst}
  &
    \text{If $\Updelta; \Gamma \proves \varphi$, $B :: \sigma$,
      and ${\bar x}_\sigma \notmetain \FV_\set(\Updelta \Cup \Gamma)$,
      then $\Updelta; \Gamma \proves \varphi[{\bar x}_\sigma := B]$}
  \\
    \mathlabel{ext}
  &
    \text{If $\Updelta; \Gamma \proves \app F {\bar x}_\sigma =_\tau \app G {\bar x}_\sigma$,
      then $\Updelta; \Gamma \proves F =_{\sigma \Fun \tau} G$}
\end{array}}

\newcommand{\ConstTypes}{\begin{array}{@{}c@{}}
    {\imp}, {\iff}, \land, \lor \ccolon \bool \Fun \bool \Fun \bool;
  \qquad
    {\neg} \ccolon \bool \Fun \bool;
  \qquad
    \True, \False \ccolon \bool;
  \qquad
    {=} \ccolon \alpha \Fun \alpha \Fun \bool;
  \\[.2ex]
    \forall, \exists, \exOneSym, \atMostOneSym
               \ccolon (\alpha \Fun \bool) \Fun \bool;
  \qquad
    \riota \ccolon \alpha \Fun (\alpha \Fun \bool) \Fun \alpha;
  \qquad
    \ifconst \ccolon \bool \Fun \alpha \Fun \alpha \Fun \alpha
  \end{array}}
\newcommand{\True}{\mathsf{True}}
\newcommand{\False}{\mathsf{False}}
\newcommand{\ZFCinHOL}{\mathsf{ZFCinHOL}}
\newcommand{\HOL}{\mathsf{HOL}}
\newcommand{\Ax}{\mathsf{Ax}}

\section{Mathematical Definitions and Logical Framework}
\label{sec:math-defin-logic}

\def\bbslash{\diediediediediedie}

We assume a set-theoretic meta-level with at least:
  equality $\meq$;
  definitions $\abbreviates$;
  an empty set $\varnothing$;
  set membership $\metain$,
  binary union $\Cup$,
  set literals $\mset{ \square_1, \ldots, \square_n }$,
  and comprehensions $\mset{ \square \mid \square }$;
  an empty list $\nil$,
  tuple/list literals $[\square_1, \ldots, \square_n]$,
  membership $\listin$,
  cons $\cons$,
  append $\append$,
  and comprehensions $\mlist{ \square \mid \square }$;
and
  a set of natural numbers $\nat$.
Uses of ${\elem}$ declare that symbols
  on the left side of ${\elem}$ are metavariables ranging over the set on the right side.

\subsection{Syntax and Types}

\begin{boxfigure}[t]{fig:base-syntax}
  {Base syntax, type variables, constants, terms, and the typing relation.}
  $$\predisplaysize=-9999pt\relax
    \BaseSyntax$$
  \hrule
  $$\predisplaysize=0pt\relax
    \FreeTypeVariables$$
  \hrule
  $$\predisplaysize=0pt\relax
    \ConstTypes
  $$
  \hrule
  $$\predisplaysize=0pt\relax
    \TermSyntax$$
  \hrule
  $$
    \predisplaysize=-99pt\relax
    \belowdisplayshortskip=0pt\relax
      \vTypeRule
    \qquad
      \cTypeRule
    \qquad
      \lamTypeRule
    \qquad
      \appTypeRule
  $$
\end{boxfigure}

Our logical framework is close to and inspired by
   the Isabelle/HOL formulation of Kun\v{c}ar and Popescu~\cite{kuncar2019}.
\Autoref{fig:base-syntax} defines the meta-level sets
  $\TVar$ of \emph{type variables},
  $\RVar$ of \emph{raw term variables},
  and $\Type$, $\Var$, and $\Const$.
$\Type$ consists of type variables,
  the type $\bool$ of truth claims/assumptions,
  \emph{domain types} $\domain i$,
and \emph{operator types} $\sigma \Fun \tau$.%
\footnote{%
  At object level, ``function'' means an object satisfying the $\Func$ predicate of
  \autoref{fig:cla-defs2}.}
The constructor $\Fun$ is right associative so
$(\tau_1 \Fun \tau_2 \Fun \tau_3)
\meq (\tau_1 \Fun (\tau_2 \Fun \tau_3))$.
$\TV_\typ(\sigma)$ yields the set of type variables occurring in $\sigma$.
$\Var$ is the set of \emph{term variables}, which are pairs of raw term variables and types.
A term variable $[\bar x, \sigma]$ may be written as ${\bar x}_\sigma$.
$\Const$ is a set of \emph{constants}.
The metavariable $\mathcal{K}$ ranges over lists of constants.
The fixed meta-level function $\ctyp : \Const \to \Type$ assigns each constant a type.
We write $\bar\kappa \ccolon \tau$ for $\ctyp(\bar\kappa) \meq \tau$.
\Autoref{fig:base-syntax} assigns some constants some types.

The notation $\sigma[\alpha := \tau]$ denotes the \emph{type substitution} 
  replacing occurrences of the type variable $\alpha$ in $\sigma$ by $\tau$.
A type $\sigma$ is an \emph{instance} of $\tau$ (written $\sigma \instOf \tau$)
  iff $\sigma \meq \tau[\alpha_1 := \rho] \ldots [\alpha_n := \rho_n]$ 
  for some $\rho_1, \ldots, \rho_n$.
For example, $(\domain 1 \Fun \domain 1) \instOf (\alpha \Fun \alpha)$.
A \emph{constant instance} is a pair $[\bar\kappa, \sigma]$
  such that $\sigma \instOf \ctyp(\bar\kappa)$
(e.g., 
  $[\cPow, \domain 0 \Fun \domain 0]$ and 
  $[{\in}, \domain 1 \Fun \domain 1]$ are constant instances
  for the constants $\cPow \ccolon \alpha \Fun \alpha$
  and ${\in} \ccolon \alpha \Fun \alpha \Fun \bool$ respectively). 
Constant instances $[\bar\kappa, \sigma]$ may be written as ${\bar\kappa}_\sigma$,
  and if $\ctyp(\bar\kappa)$ has no type variables,
  then we may write $\bar\kappa$ for
  the constant instance $\bar\kappa_{\ctyp(\bar\kappa)}$.
Let $\theta,\kappa$ range over constant instances.

\Autoref{fig:base-syntax} defines the set $\RTerm$ of \emph{raw terms}.
A raw term is
  a variable $x$,
  a constant instance $\kappa$,
  an \emph{application} $\app{\bar B}{\bar C}$,
  or an \emph{abstraction} $\lam{x}{\bar B}$.
Define the \emph{free variable} meta-level function $\FV_\trm$ and $\upalpha$-equivalence on
$\RTerm$ as usual.
Let $\ATerm$ be $\RTerm$ modulo $\upalpha$-equivalence.
Define \emph{term substitution} and \emph{type substitution} on $\ATerm$ as usual with the notation
$\bar B[x\,{:=}\,\bar C]$ and $\bar B[\alpha\,{:=}\,\sigma]$.
Define $\upbeta$-equivalence on $\ATerm$ as usual.
Let $\Term$ be $\ATerm$ modulo $\upbeta$-equivalence.
Lift $\FV_\trm$ and term and type substitution to $\Term$.

Let the typing relation $::$ between $\Term$ and $\Type$ be the least
relation satisfying the rules in \autoref{fig:base-syntax}.
Let a term $B$ be a \emph{formula} iff $B :: \bool$.
A \emph{simple definition} is a formula of the form
$\kappa =_\sigma B$.%
\footnote{$B$ must not refer to $\kappa$, including via a chain of other definitions.}
Let $\varphi$, $\psi$, and $\gamma$ range over formulas,
  let $\Updelta$, $\Gamma$, and $\Uptheta$ range over sets of formulas,
  and let $\Upphi$ range over formula lists.

We adopt the following notation.
Given $x \meq {\bar x}_\sigma$, the expression
 $\lam {x :: \sigma}{B}$ stands for $\lam{\bar{x}_\sigma}{B}$.
Inside term expressions $B =_\sigma C$ and $\lam {x :: \sigma} B$,
  we allow omitting the type $\sigma$ (and the $::$),
  provided that $\sigma$ can be uniquely determined
  by the typing rules and other type information in or about $B$ and $C$.
The notation $\lam {x_1\, \cdots\, x_n}{B}$ stands for the
  nested abstractions $\lam {x_1}{\cdots\,(\lam{x_n}{B})}$.
The notation $\mlist{\kappa_1 :: \tau_1, \ldots, \kappa_n :: \tau_n}$
   denotes the list $\mlist{\kappa_1, \ldots, \kappa_n}$
   and asserts that
   $\kappa_i \metain \Const$ and $\kappa_i \ccolon \tau_i$ for $1 \leq i \leq n$.

If $\bar\kappa$ is \emph{infix},
  an application $\app{(\app{\bar\kappa_\sigma}{B})}{C}$
  is written as $B \mathrel{\bar\kappa_\sigma} C$.
The constants $=$, $\land$, $\lor$, and $\imp$ are all infix,
  and listed here in descending order of precedence.
Negation and operator application take precedence over infix operators,
  e.g., $\app \neg P \land \app \neg Q$ is $(\app \neg P) \land (\app \neg Q)$
  and $\app F x =_\sigma \app G x$ is $(\app F x) =_\sigma (\app G x)$.
If $\bar\kappa$ is a \emph{binder},
  an application $\app{\bar\kappa_\sigma}{(\lam {x :: \sigma} B)}$
  is written as $(\binddot{\bar\kappa}{x :: \sigma}{B})$,
and $\binddot{\bar\kappa}{x_1,\ldots,x_n}{B}$ stands for
  the nested applications of quantifiers and abstractions
  $\app{\bar\kappa_\sigma}
    {(\lam{x_1 :: \sigma}
      {\cdots (\app{\bar\kappa_\sigma}{(\lam{x_n :: \sigma}{B})})})}$.
The constants $\forall$, $\exists$, $\exOneSym$, and $\atMostOneSym$ are all binders.

\begin{boxfigure}[t]{fig:soft-types+inf-rules}
   {Types, definitions and notation for soft types, axioms, and inference rules.}
  $$\predisplaysize=0pt\relax
  \begin{array}{@{}r@{\:\:\ccolon\:\:}l@{\quad}r@{\:\:\ccolon\:\:}l@{}}
      ({:})
    &
      \alpha \Fun (\alpha \Fun \bool) \Fun \bool
  &
      {\fun}
    &
      (\alpha \Fun \bool) \Fun (\beta \Fun \bool) \Fun
        \left((\alpha \Fun \beta) \Fun \bool \right)
  \\
      \top, \bot
    &
      \alpha \Fun \bool
  &
     \Uppi
    &
      (\alpha \Fun \bool) \Fun (\alpha \Fun \beta \Fun \bool) \Fun
        \left((\alpha \Fun \beta) \Fun \bool \right)
  \\
      {\subtyp}
    &
      (\alpha \Fun \bool) \Fun (\alpha \Fun \bool) \Fun \bool
  &
      {\inter}, {\uni}
    &
      (\alpha \Fun \bool) \Fun (\alpha \Fun \bool) \Fun (\alpha \Fun \bool)
  \end{array}
  $$
  \hrule
  \quad
  $
    \begin{array}{@{}l@{}}
        \SoftTypes
      \abbreviates
        \llbrace
        \ 
            ({:})
          =
            \lam{x\, p}{\app p x}
        ,
        \\\qquad\quad
            \top
          =
            \lam{x}{\True}
        ,
          \quad
            \bot
          =
            \lam{x}{\False}
        ,
        \\\qquad\quad
            {\fun}
          =
            \lam{p\,q\,f}{\all x {(x : p \imp \app f x : q)}}
        ,
        \\\qquad\quad
            {\Uppi}
          =
            \lam{p\,q\,f}{\all x {(x : p \imp \app f x : \app q x)}}
        ,
        \\\qquad\quad
            {\inter}
          =
            \lam{p\,q\,x}{(x : p \land x : q)}
        ,
        \\\qquad\quad
            {\uni}
          =
            \lam{p\,q\,x}{(x : p \lor x : q)}
        ,
        \\\qquad\quad
            {\subtyp}
          =
            \lam{p\,q}{\all x {(x : p \imp x : q)}}
        \ 
        \rrbrace
    \end{array}
  $
  \quad
  \vrule
  \quad
  $\begin{array}{l@{\:\abbreviates\:}l}
      \depfun{x : P}{Q}
    &
      \apptwo\Uppi P {(\lam x Q)}
  \\
      \all {x : P} {\varphi}
    &
      \all x {(x : P \imp \varphi)}
  \\
      \ex {x : P} {\varphi}
    &
      \ex x {(x : P \land \varphi)}
  \\
      \atMostOne {x : P} {\varphi}
    &
      \atMostOne x {(x : P \land \varphi)}
  \\
      \exOne {x : P} {\varphi}
    &
      \exOne x {(x : P \land \varphi)}
  \end{array}$
\vskip 0.5em\relax
\hrule
  $\begin{array}{@{\quad}r@{\:}l}
    \HOL & \abbreviates
    
    \llbrace\: 
      \all x {x =_{\alpha} x},\quad
      \all {p,x,y} {x =_{\alpha} y \imp \app px \imp \app py},\quad
      \all p {p = \True \lor p = \False},
    \\ & \qquad 
      \all {p,d} {{(\app\exOneSym p)}\imp \app p{(\apptwo\riota dp)}},\quad
      \all {p,d} {(\app\neg{\app\exOneSym p}) \imp {\apptwo\riota dp}=d},
    \\ & \qquad  
      \ifconst=\lam{b\,x\,y}
      \defdes c {(b\imp c=x)\land(\app\neg b\imp c=y)\,} {\,x},

      \:\ldots\:
    \rrbrace
  \end{array}$  
\vspace{0.5em}
\hrule
  $$\predisplaysize=-9999pt\relax
    \belowdisplayshortskip=0pt\relax
    \InferenceRules$$
\end{boxfigure}

\subsection{Soft Types}\label{sec:soft-types}

A \emph{soft type} is an operator of type $\tau\Fun\bool$ for some
$\tau$.
A key difference from ``hard'' types is that each object will satisfy many (often infinitely many)
soft types.
\Autoref{fig:soft-types+inf-rules} gives types and simple definitions for operators for building and
using soft types.
Our soft type constructors are mostly the same as those used by Kappelmann, Chen, and Krauss in
Isabelle/Set~\cite{Kap+Che+Kra:Isabelle-Set-git-2022-05}, which in turn are along the lines
suggested much earlier by Krauss~\cite{krauss2010}.
To help the reader think ``types'', and also to allow proof tactics to follow soft-type-specific
strategies, we write ``$F : P$'' as a \emph{soft typing} which means the same thing as ``$\app
PF$'', i.e., $F$ satisfies the predicate $P$.
We also have a non-dependent constructor ${\fun}$ and a dependent constructor $\Uppi$ for soft types
on operators,
  soft intersection and union type constructors ${\inter}$ and ${\uni}$,
  and soft subtyping ${\subtyp}$.
  Our development derives the standard introduction and elimination rules
  for each of these concepts.
  The constants
  ${(:)}$, ${\fun}$, ${\inter}$, ${\uni}$,
  and ${\subtyp}$ are all infix,
  and $\fun$ is right-associative.
  The figure gives notation for dependent operator soft types,
  restricted quantification,
  and restricted binding of $\lambda$-expressions.
Let $\SoftTypes$ be the set of simple definitions as defined in
\autoref{fig:soft-types+inf-rules}.

For example, using the soft operator type constructor $\fun$, given a predicate
$P::\sigma\Fun\bool$ and a predicate $Q::\tau\Fun\bool$, the term $P\fun Q$ is a predicate of type
$(\sigma\Fun\tau)\Fun\bool$ such that if $P\fun Q$ is true of $x::\sigma\Fun\tau$ and $P$ is true of
$y::\sigma$ then $Q$ is true of $\app x y::\tau$.
Precisely, $F : P\fun Q$ means $\all b {\,b : P \imp \app F b : Q}$.

\subsection{Inference Rules and Axioms}

Let $\Gamma + \varphi$ denote $\Gamma \Cup \mset{\varphi}$.
Let $\TV_\trm$ be the extension of $\TV_\typ$ to terms.
Let $\FV_\st(\Gamma)$ be the union of all $\FV_\trm(\varphi)$ for all $\varphi \metain \Gamma$.
Let $\TV_\st(\Gamma)$ be the union of all $\TV_\trm(\varphi)$ for all $\varphi \metain \Gamma$.
Let $\Gamma[x := B]$ be the set of all $\varphi[x := B]$ for all $\varphi \metain \Gamma$.

$\HOL$ is the set of axioms (formulas) given in \autoref{fig:soft-types+inf-rules}
that implement reflexivity of equality,
indiscernability of equal objects,
the law of the excluded middle,
a definite description operator with a default ($\riota$),
a conditional operator ($\ifconst$),
and simple definitions (not shown) for
$\True$, $\forall$, $\exists$, $\False$,
$\neg$, ${\land}$, ${\lor}$, $\iff$, $\atMostOneSym$, and $\exOneSym$.
Nice notation for $\riota$ and $\ifconst$ are given thus:
  $(\defdes x \varphi D) \abbreviates \apptwo \riota D{(\lam x \varphi)}$,
  $(\ifelse P B C) \abbreviates \appthree \ifconst P B C$.
Let $\ZFCinHOL$ be the set of axioms used in Paulson's ``ZFC in HOL''~\cite{Pau:ZFC_in_HOL-AFP:2019}.
Let the \emph{deduction relation} $\proves$ be the least relation satisfying
the inference rules in \autoref{fig:soft-types+inf-rules}.
Our normal use will be to derive judgements of the form
$\HOL\Cup\SoftTypes\Cup\Updelta; \Gamma \proves \varphi$
where $\Updelta$ contains additional definitions specific to the topic of the proof and $\Gamma$
contains local assumptions.

\subsection{Classes}

A \emph{(type) class} is a tuple 
  $\cla \meq [\cladeps, \claconsts, \Upphi, \Uptheta]$.
$\cladeps$ is a list of classes called the \emph{dependencies} of $\cla$.
$\claconsts$ is a list $[\kappa_1, \ldots, \kappa_n]$ 
  of pairwise distinct constants called the \emph{parameters} of $\cla$, 
  such that 
  $\TV_\trm(\kappa_1) \Cup \ldots \Cup \TV_\trm(\kappa_n) \meq \mset{\alpha}$ 
  for some $\alpha$,
i.e., exactly one type variable
      occurs in the parameters of $\cla$,
      which we refer to as $\tv(\cla)$.
  $\Upphi$ is a list of formulas
and
  $\Uptheta$ is a list of simple definitions,
called the \emph{axioms} and \emph{definitions} of $\cla$ respectively.

We write $\sigma\typSub{\domain i}$ and $B\typSub{\domain i}$
  for the results of the type substitutions
    $\sigma[\tv(\cla) := \domain i]$ and 
    $B[\tv(\cla) := \domain i]$ respectively.
A \emph{parameter instantiation} of $\cla$ at $\domain i$ is a set of formulas
   $\mset{\kappa_1\typSub{\domain i} = B_1, \ldots,
          \kappa_n\typSub{\domain i} = B_n}$,
  where $\TV(B_j) \meq \varnothing$ for $j\metain\mset{1,\ldots,n}$.
Relative to a set of hypothesis $\Gamma$ and a set of definitions $\Updelta$,
  we say that $\cla$ is \emph{instantiated} at $\domain i$
  if we have $\Updelta; \Gamma \proves \varphi$
  for any $\varphi \in \Upphi\typSub{\domain i}$.
Typically $\Updelta$ will contain parameter instantiations 
  for all of the dependencies of $\cla$.

\section{GSTs as Type Classes}
\label{sec:gsts-as-type}

\subsection{GST Features}
\label{sec:gst-features}

\begin{boxfigure}[t!]{fig:cla-defs}%
  {Constants, axioms, and definitions for the $\GZF$, $\Ordinal$, and $\OrdinalRec$ classes}%
  $$%
    \predisplaysize=-9999pt\relax
    \belowdisplayshortskip=0pt\relax
    \begin{array}{rll}
      \GZF_\consts & \abbreviates [
    &
      \defaultval\GZF :: \alpha,\:
      \Set :: \alpha \Fun \bool,\:
      {\in} :: \alpha \Fun \alpha \Fun \bool,\:
      \cUnion :: \alpha \Fun \alpha,\:
      \cPow :: \alpha \Fun \alpha,\:
  \\ & &
      \Emp :: \alpha,\:
      \cSucc :: \alpha \Fun \alpha,\:
      \Inf :: \alpha,\:
      \cRepl :: \alpha \Fun (\alpha \Fun \alpha \Fun \bool) \Fun \alpha
    \:]
  \\
      \GZF_\axioms & \abbreviates [
    &
      \cUnion : \SetOf \Set \fun \Set,
    \quad
      \cPow : \Set \fun \SetOf \Set,
    \\ & &
      \Emp : \Set,
    \quad
      \cSucc : \Set \fun \Set,
    \quad
      \Inf : \Set,
    \\ & &
      \cRepl : (\depfun {x : \Set} {\ReplPred x \fun \Set}),
    \\ & &
      \all {x,y : \Set} {\all {b} {(b \in x \iff b \in y) \imp x = y}},
    \\ & &
      \all {x : \SetOf \Set}
        {\all b {b \in \Union x \iff (\ex y {y \in x \land b \in y})}},
    \\ & &
      \all {x,y : \Set} {y \in \Pow x \iff y \subseteq x},
    \\ & &
      \all b {\app \neg {b \in \Emp}},
    \quad
      \all {x : \Set} {b \in \Succ x \iff (b \in x \lor b = x)},
    \\ & &
      \Emp \in \Inf \land (\all b {b \in \Inf \imp \Succ b \in \Inf}),
    \\ & &
      \all {x : \Set} {\all {p : \ReplPred x}
    \\ & & \quad
        {\all c {c \in \Repl x p \iff
          (\ex b {b \in x \land \apptwo p b c \land c : \SetMem})}}}
    \:]
    \\
    \GZF_\defs & \abbreviates [
    &
      {\subseteq} = (\lam {x\,y} {\all b {b \in x \imp b \in y}}),
    \quad
      \SetMem = (\lam b {\ex {y : \Set} {b \in y}}),
    \\ & &
      \SetOf = (\lam {p\, x} {x : \Set \land {\all {b \in x}{b : p}}}),
    \\ & &
      \cReplPred  = (\lam {x\, p}
        {\all {b \in x}{\atMostOne {c : \SetMem} {\apptwo p b c}}})
    \:]
    \\[2ex]
      \hline
    \\
      \Ord_\consts & \abbreviates
      [ &
        \defaultval\Ord :: \alpha,\,
        \Ord :: \alpha \Fun \bool,\,
        {<} :: \alpha \Fun \alpha \Fun \bool,\,
        0 :: \alpha,\,
        \csucc :: \alpha \Fun \alpha,\,
        \omega :: \alpha
      \,]
    \\
      \Ord_\axioms & \abbreviates
    [ &
      0 : \Ord,
    \quad
      \csucc : \Ord \fun \Ord,
    \quad
      \omega : \Limit,
    \\ & &
      \all {u : \Ord} {\app \neg {u < 0}},
    \\ & &
      \all {u,v : \Ord} {u < \succ v \iff (u < v \lor u = v)},
    \\ & &
      \all {u : \Limit} {u = \omega \lor \omega < u},
    \\ & &
      \all {u,v,w : \Ord} {u < v \imp v < w \imp u < w}\quad\mbox{(transitivity)},
    \\ & &
      \all {u,v : \Ord} {u < v \imp \app \neg {v < u}}\quad\mbox{(antisymmetry)},
    \\ & &
      \all {u, v : \Ord} {u < v \lor u = v \lor v < u}\quad\mbox{(trichotomy)},
    \\ & &
      \all p {(\all {u : \Ord} {(\all {v : \Ord} {v < u \imp \app p v}) \imp \app p u})
         \imp (\all {w : \Ord} {\app p w})}
    \:]
  \\
    \Ord_\defs & \abbreviates
  [ &
    \Limit =
      (\lam u {u : \Ord \land 0 < u
        \land (\all {v : \Ord} {v < u \imp \succ v < u})})
  \:]
\\[2ex]\hline
\\
  \cOrdRec_\consts & \abbreviates
    [ &
      \defaultval\cOrdRec :: \alpha,\:
      \cpredSet :: \alpha \Fun \alpha,\:
      \csupOrd :: \alpha \Fun \alpha,\:
    \\ & &
      \cOrdRec ::
        (\alpha \Fun (\alpha \Fun \alpha) \Fun \alpha) \Fun
        (\alpha \Fun \alpha \Fun \alpha) \Fun
        \alpha \Fun \alpha \Fun \alpha
    \,]
  \\
    \cOrdRec_\axioms & \abbreviates
  [ &
    \cpredSet : \Ord \fun \SetOf \Ord,
  \quad
    \csupOrd : \SetOf \Ord \fun \Ord,
  \\ & &
    \all {u,v : \Ord} {u \in \predSet v \iff u < v},
  \\ & &
    \all {x : \SetOf \Ord} {\all u {u \in x \imp u < \succ{(\supOrd x)}}},
  \\ & &
    \all {g, f, x} {\OrdRec g f x {0} = x},
  \\ & &
    \all {g, f, x} {\all {u : \Ord}
      {\OrdRec g f x {(\succ u)} = \apptwo f {(\succ u)} {(\OrdRec g f x u)}}},
  \\ & &
    \all {g, f, x} {\all {u : \Limit}
      {\OrdRec g f x u =
  \\ & &
    \qquad
        \apptwo g u
          {(\lam v {\ifelse {v : \Ord \land v < u} {\OrdRec g f x v} {\defaultval\cOrdRec}})}}}
  \:]
  \end{array}$$%
\end{boxfigure}

\begin{boxfigure}[t]{fig:cla-defs2}
  {Constants, axioms, and definitions for the $\Function$ class}
  $$
    \predisplaysize=-9999pt\relax
    \belowdisplayshortskip=0pt\relax
  \begin{array}{r@{\,}l}
    \Func_\consts \abbreviates
    [ &
      \defaultval\Func :: \alpha,\:
      \Func :: \alpha \Fun \bool,\:
      \fapp :: \alpha \Fun \alpha \Fun \alpha \Fun \bool,\:
      {\pfun} :: \alpha \Fun \alpha \Fun \alpha,
    \\ &
      \cmkfun :: \alpha \Fun (\alpha \Fun \alpha \Fun \bool) \Fun \alpha,\:
      \cdom :: \alpha \Fun \alpha,\:
      \cran :: \alpha \Fun \alpha
    \,]
  \\
    \Func_\axioms \abbreviates
  [ &
    \cmkfun : (\depfun {x : \Set} {\FunPred x \fun \Func}),
  \\ &
    \cdom : \Func \fun \Set,\:
    \cran : \Func \fun \Set,
  \\ &
    {\pfun} : \Set \fun \Set \fun \SetOf \Func,
  \\ &
    \all {f : \Func} {\all {b,c,d}
      {\fapp f b c \land \fapp f b d \imp c = d}}
  \\ &
    \all {f,g : \Func}
      {(\all {b,c} {\fapp f b c \iff \fapp g b c}) \imp f = g}
  \\ &
    \all {f : \Func}
      \all b {b \in \dom f \iff (\ex c {\fapp f b c})}
  \\ &
    \all {f : \Func}
      \all c {c \in \ran f \iff (\ex b {\fapp f b c})}
  \\ &
    \all {x,y : \Set} {\all {f : \Func}
      {(f \in x \pfun y) \iff (\dom f \subseteq x \land \ran f \subseteq y)}}
  \\ &
    \all {x : \Set} {\all {p : \FunPred x}
    \all {b,c}
  \\ &
      {\quad\fapp {(\mkfun x p)} b c \iff
        (b \in x \land \apptwo p b c \land b : \FunMem \land c : \FunMem)}}
  \:]
  \\
  \Func_\defs \abbreviates
  [ &
    \FunMem = (\lam {b} {\ex {f : \Func}
                {b \in \dom f \lor b \in \ran f}}),
  \\ &
    \FunPred = (\lam {x\, p}
          {\all {b : \FunMem} {b \in x \imp (\atMostOne {c : \FunMem} {\apptwo p b c})}})
  \:]
  \end{array}$$
\end{boxfigure}

\begin{boxfigure}[t!]{fig:cla-features}
  {Definition of
   our example
   features and their classes.}
  $$
    \begin{array}{rcl}
      \GZF & \abbreviates &
        \mlist{\nil,\GZF_\consts, \GZF_\axioms, \GZF_\defs}
    \\
      \Ordinal & \abbreviates &
        \mlist{\nil, \Ord_\consts, \Ord_\axioms, \Ord_\defs}
    \\
      \OrdinalRec & \abbreviates &
        \mlist{\mlist{\GZF, \Ordinal},
          \cOrdRec_\consts, \cOrdRec_\axioms, \cOrdRec_\defs}
    \\
      \Function & \abbreviates &
        \mlist{\mlist{\GZF}, \Func_\consts, \Func_\axioms,\Func_\defs}
    \\
      \Exception & \abbreviates &
        \mlist{\nil,
          \mlist{\circ_\Exc :: \alpha, \Exc :: \alpha \Fun \bool, \bullet :: \alpha}, \nil, \nil}
    \end{array}
  $$
  \hrule
  $$
    \predisplaysize=-9999pt\relax
    \abovedisplayshortskip=0pt\relax
    \belowdisplayshortskip=0pt\relax
  \begin{array}{r@{\:\abbreviates\:}l}
    \bfGZF &
      \mlist{\GZF, \Set, \SetMem, \defaultval\GZF}
  \\
    \bfOrdinal &
      \mlist{\Ordinal, \Ord, \bot, \defaultval\Ord}
  \\
    \bfOrdRec &
      \mlist{\OrdinalRec, \bot, \bot, \defaultval\cOrdRec}
  \\
    \bfFunction &
      \mlist{\Function, \Func, \FunMem, \defaultval\Func}
  \\
    \bfExc &
      \mlist{\Exception, \Exc, \bot, \defaultval\Exc}
  \end{array}
  $$
\end{boxfigure}

A \emph{feature} is a tuple 
  $\feat \meq [\,\cla, P_\logo, P_\cargo, \kappa_\default\,]$,
  where $\cla$ is a class,
  $P_\logo, P_\cargo :: \tv(\cla) \Fun \bool$
  are the \emph{cargo} and \emph{logo} soft types of $\feat$,
  and $\kappa$ is a constant called the \emph{default parameter} of $\feat$.
The terms $P_\logo, P_\cargo, \kappa_\default$
  keep track of information used when combining features to create GSTs.
$P_\logo$ (a.k.a.\ $\logo(\feat)$)
  should be chosen as the soft type of all objects contributed to the
  domain by a feature's axioms.
For features that do not contribute any objects to the domain,
 $\bot$ should be used.
$P_\cargo$ (a.k.a.\ $\cargo(\feat)$)
  should be chosen as the soft type satisfied of all objects
  contained in the internal structure of some object $X : P_\logo$.
Keeping track of cargo types allows preventing the exception object $\bullet$
  from being contained in sets and functions, allowing the benefits of a free logic,
  i.e., terms can be ``undefined'',
  without the need for anything to actually really be undefined and without the need for separate
  quantifiers for an ``inner'' and ``outer'' domain.
  The $\bullet$ object is in the spirit of concepts like the number $0$ and the empty set $\Emp$,
  i.e., it is a object representing what would otherwise be the lack of an object.
Each feature's class has its default parameter $\kappa_\default$
  (a.k.a.\ $\default(\feat)$) in its list of parameters,
  given as the symbol $\circ$ decorated with a subscript.
The default parameter is intended to be a placeholder that can be used in a feature's axioms and
  definitions for exceptional results and normally it will be axiomatized to be equal to some
  specific object (typically $\bullet$) when features are combined into a GST.

\subsection{Example Features}\label{sub:example-features}

We now define features for
  sets, ordinals, ordinal recursion,
  ordered pairs, functions, and an exception ($\bullet$).
Figures~\ref{fig:cla-defs} and~\ref{fig:cla-defs2} define
   constants, axioms, and definitions
  for each feature's class.
\Autoref{fig:cla-features} defines the classes
  and their features (in boldface).
The Isabelle/HOL development can be found in
  \texttt{GST\_Features.thy}.

We have formulated some of our features' axioms as soft typings.
This is more useful in a GST than in set-only ZF, because even basic set operations like
$\cUnion$ and $\cPow$ can be ``undefined'' because, e.g., the argument might be a non-set

We use a number of soft types that classify objects in a GST's domain, e.g., $\Set$,
$\Function$, and $\Pair$.
For example, $\Emp:\Set$.
An example of using the soft operator type constructor $\fun$ is that using the axioms
$\cdom:\Func\fun\Set$ and and $\cPow:\Set\fun\SetOf\Set$ and $\cUnion:\SetOf\Set\fun\Set$, we can
deduce that if $f$ is a function (i.e., $f:\Func$), then
$T=\app{\cUnion}{(\app{\cPow}{(\app{\cdom}{f})})}$ is a set (i.e., $T:\Set$) and is not
``undefined'' (i.e., $T\ne\bullet$).
Our soft types are similar in spirit to the soft types of Mizar~\cite{wiedijk2007-mizar}, but differ
in many details.

We define specialized soft type constructors for our features, e.g., using the $\cSetOf$ soft type
constructor we can build the soft type $\SetOf\Set$ of those sets that contain only sets.
For example, $\{0\}:\SetOf{\Ord}$, because $\{0\}$ is a set containing only ordinals, and
$\{\{0\}\}:\SetOf{\Set}$, and also
$\{\{0\}\}:\SetOf{(\SetOf{\Ord})}$.

The feature $\bfGZF$ provides \emph{generalized Zermelo/Fraenkel} sets.
$\GZF_\defs$ defines the cargo soft type $\SetMem$,
   a soft type for objects that belong to some set in the domain.
$\bfOrdinal$ provides ordinal numbers.
$\bfOrdRec$ provides an operator for recursion on ordinals,
  which is crucial for building models of GSTs.
Adding the $\bfOrdinal$ and $\bfOrdRec$ features saved us at least a month of development time
because it allows us to pass the development of ordinal recursion from a type implementing
\texttt{ZFC\_in\_HOL} to a GST whose model is built within that type.
$\bfFunction$ provides functions
  with a function application relation $\capp$,
  operators to find the domain and range of a function, a partial function space operator
  ${\pfun}$,
  and a function-building operator $\cmkfun$.
$\bfExc$ provides the \emph{exception object} $\bullet$ (spoken ``boom'').
The important behavior of $\bullet$ is given by axioms generated when combining features
that use cargo soft types (e.g., $\SetMem$, $\FunMem$) to ensure that $\bullet$ can not
occur within container objects (e.g., sets, functions).

Our current design requires each object in a GST's domain to be ``owned'' by exactly one
feature.
However, classes associated with a feature may be used independently of that feature and can be
instantiated by any GST's domain provided some collections of objects
in the domain can be found that satisfy the requirements the class places on parts of the type.
For example, $\Function$ can be instantiated by sets of ordered pairs (cf.\ \texttt{GZF/SetRel.thy}),
and we do this to build a GST at type $\mathbf{V}$ (our founder domain).

\subsection{Feature Combination}

\label{sec:combination}

\begin{boxfigure}[t]{fig:feature-axs}
  {`Otherwise' axioms for operators, logo axioms, and cargo axioms.}
  $$
\begin{array}{c}
\begin{array}{l@{\:\abbreviates\:}l}
  \typList_i(P \fun Q) &
    b_i : P \cons \typList_{i+1}(Q)
\\
  \typList_i(\apptwo\Uppi {P} {Q}) &
    {b_i} : P \cons \typList_{i+1}(\app Q {b_i})
\\
  \typList_i(R) & \nil
\end{array}
\\[5mm]
\begin{array}{l}
  \otherwise(\kappa, [b_1 : P_1, \ldots, b_n : P_n], D)
  \abbreviates \\ \qquad
    \all {b_1, \ldots, b_n}
      {(\lnot b_1 : P_1 \lor \ldots \lor \lnot b_n : P_n)
        \imp (\appldots\kappa{b_1}{b_n} = D)}
\end{array}
\\[3mm]
\begin{array}{l@{\:}c@{\:}l}
    \allOtherwise(\feat, D)
  & \abbreviates &
    [\, \otherwise(\kappa, \typList(R), D)
      \mid (\kappa : R) \listin \axioms(\feat),
  \\ & & \qquad\qquad
      R \meq (P \fun Q) \text{ or }
      R \meq (\depfun {x : P} Q) \,]
\end{array}

\end{array}
$$
\hrule
$$
  \predisplaysize=0pt\relax
\begin{array}{r@{\:}l}
  \cover([P_1, \ldots, P_n]) \abbreviates &
    (P_1 \uni \ldots \uni P_n = \top)
\\
  \disjoint([P_1, \ldots, P_n]) \abbreviates &
    \mlist{(P_1 \inter P_2 = \bot), \ldots, (P_1 \inter P_n = \bot),
           \ldots, (P_{n-1} \inter P_n = \bot)}
\end{array}
$$
\hrule
$$
  \predisplaysize=0pt\relax
  \belowdisplayshortskip=0pt\relax
\begin{array}{r@{\:}c@{\:}l}
  \admitCargo(P, \mlist{Q_1, \ldots, Q_n})
  & \abbreviates &
    (Q_1 \uni \ldots \uni Q_n \subtyp P)
\\
  \restrictCargo(P, \mlist{Q_1, \ldots, Q_n})
  & \abbreviates &
    ((Q_1 \uni \ldots \uni Q_n) \inter P = \bot)
\\
  \cargoAx(\feat, \wlist, \blist)
  & \abbreviates &
    [\:\admitCargo
      (\cargo(\feat),
       \mlist{ \logo(\featG)
        \mid \featG \listin \wlist, \featG \not\listin \blist}),
\\
  &  &
    \:\: \restrictCargo
      (\cargo(\feat),
       \mlist{ \logo(\featG)
        \mid \featG \listin \blist}) \:]
  \end{array}
$$
\end{boxfigure}

\newcommand{\spec}{\mathsf{spec}}
\begin{boxfigure}[t]{fig:mkgst}
  {Generating GSTs from specifications and $\ZFplus$, an example GST.}
  Given $\spec \meq [[\feat_1, D_1, \blist_1], \ldots, [\feat_n, D_n, \blist_n]]$, then:
  $$
  \begin{array}{r@{\:}c@{\:}l}
    \GST_\axioms(\spec) & \abbreviates
    &
      \allOtherwise(\feat_1, D_1)
      \append \ldots \append
      \allOtherwise(\feat_n, D_n)
    \\ & \append &
      \app{\disjoint}{\mlist{\logo(\feat_1), \ldots, \logo(\feat_n)}}
    \\ & \append &
      \mlist{\app\cover{\mlist{\logo(\feat_1), \ldots, \logo(\feat_n)}}}
    \\ & \append &
      \cargoAx
        (\feat_1, \mlist{\feat_1, \ldots, \feat_n}, \blist_1)
    \\ & \append & \ldots \append
      \cargoAx
        (\feat_n, \mlist{\feat_1, \ldots, \feat_n}, \blist_n)
  \\
      \GST_\defs(\spec) & \abbreviates
    &
      \mlist{ \default(\feat_1) = D_1, \ldots,
             \default(\feat_n) = D_n }
  \\[3mm]
      \mkGST(\spec) & \abbreviates
    &
      \mlist{
         \mlist{\class(\feat_1), \ldots, \class(\feat_n)}, \nil,
         \GST_\axioms(\spec), \GST_\defs(\spec)}
  \end{array}
$$
\hrule
$$
  \predisplaysize=0pt\relax
\begin{array}{r@{\:}c@{\:}l}
    \ZFplus_\spec & \abbreviates [
  &
      \mlist{\bfGZF, \bullet, \mlist{\bfExc}},
      \mlist{\bfOrdinal, \exc, \nil},
      \mlist{\bfFunction, \bullet, \mlist{\bfExc}},
      \mlist{\bfExc, \bullet, \nil}
  \:]
\end{array}
$$
$$\predisplaysize=-9999pt\relax
  \belowdisplayshortskip=0pt\relax
  \ZFplus \abbreviates
  \mkGST(\ZFplus_\spec)
$$
\end{boxfigure}

A \emph{feature configuration} is a triple $[\feat, D, \blist]$,
  where $D$ is a term to be identified with $\default(\feat)$,
  and $\blist$ is a \emph{blacklist} of features whose objects
  will be excluded from the internal structure of objects
  of feature $\feat$.
GSTs are classes defined by combining feature configurations.
\Autoref{fig:feature-axs} defines operations that generate extra axioms.

Most of our features use soft typing axioms constraining operator behavior, e.g., $\dom:\Func\fun\Set$
specifies that $\dom$ yields a set when applied to a function.
These soft typing judgements have a special status: they are also used to generate axioms
specifying what operators do in other cases, e.g., specifying that $\dom B$ will yield $\bullet$
when $B$ is not a function.
The formula
   $\otherwise(\kappa, \mlist{b_1 : P_1, \ldots, b_n : P_n}, D)$
  produces an axiom that identifies $D$ with the result of
  applying $\kappa$ to arguments that do not satisfy
  at least one of the $P_1, \ldots, P_n$.
The formula list $\allOtherwise(\feat, D)$
  calls $\otherwise$ on each parameter $\kappa$
  with an axiom in $\feat$ of the form
   $\kappa : P \fun Q$ or $\kappa : (\depfun{x : P}{Q})$.
Hence, $\allOtherwise(\bfFunction, \bullet)$ generates the axiom
    $\all {b} {\lnot b : \Func \imp \dom b = \bullet}$, among others.
We expect some users might prefer other policies and we aim to make this more flexible.

The formula $\cover{(\mlist{P_1,\ldots,P_n})}$ ensures that
   $B : P_i$ holds for some $P_i$,
and $\disjoint(\mlist{P_1,\ldots,P_n})$ ensures that
both $B : P_i$ and $B : P_j$ cannot hold for $i \neq j$.
The formula
  $\admitCargo(P, \mlist{Q_1, \ldots, Q_n})$
  states that $Q_1, \ldots, Q_n$ are all subtypes of $P$, and
  $\restrictCargo(P, \mlist{Q_1, \ldots, Q_m})$
  states that if $B : Q_i$ for any $i \leq m$, then $\lnot B : P$.
The formulas of $\cargoAx$ constrain the internal structure of the objects of a feature,
e.g., $\cargoAx(\bfGZF, \mlist{\bfGZF, \bfExc}, \mlist{\bfExc})$
   stands for:
$
  \mlist{(\Set \subtyp \SetMem), (\Exc \inter \SetMem = \bot)}
$.

\Autoref{fig:mkgst} defines $\mkGST$ and gives an example GST named $\ZFplus$.
The operation $\mkGST(\spec)$ defines a class $\cla$ 
  from a list $\spec$ of feature configurations.
The dependencies of $\cla$ are the classes of the features in $\spec$, the axioms of $\cla$ are
given by $\GST_\axioms(\spec)$, and the definitions of $\cla$ are a list of simple definitions of
the default parameter of each feature in $\spec$ given by $\GST_\defs(\spec)$.
Our example GST $\ZFplus$ has sets, non-set ordinals, non-set functions, and
a distinguished non-set $\bullet$.
Each feature is configured to use $\bullet$ as its default value,
  and $\bullet$ is blacklisted from the internal structure of sets, and functions.

\section{Examples of Working in a GST}
\label{sec:examples-working-gst}

These examples assume parameter instantiations of the class $\ZFplus$ and all the classes in its
dependencies (i.e., $\GZF$, $\Ordinal$, $\Function$, and $\Exception$) at the type $\domain 0$.
This means that from the class $\ZFplus$ and each class in its dependencies deductions can use
definitions of the constants in the class parameters as well as the class definitions and the class
axioms.
When we say ``define $X$'', we mean ``add the indicated simple definition for $X$ to the definitions
used in deductions''.

\medskip
\noindent\textbf{Ordinal Left-subtraction.}
Suppose we already have an infix ordinal addition operator
  such that if $i,j : \Ord$ then $i + j : \Ord$,
  and otherwise $i + j = \bullet$.
Define the \emph{left-subtraction}
operator $\leftsub::\domain 0\Fun\domain 0\Fun\domain 0$ on ordinals:%
$$%
    {\leftsub}
  =
    (\lam {i\,j} {\defdes k {i = j + k} \bullet})
$$%
Given ordinals $i,j : \Ord$ where $j<i$ or $j=i$, the term $i \leftsub j$ is the unique ordinal $k$
such that $i = j + k$.
If $i < j$ or if either $i$ or $j$ are not ordinals, then $i \leftsub j=\bullet$.
For example:
$$%
  \begin{array}{@{}c@{}}
      5 \leftsub 3
    =
      (\defdes k {5 = 3 + k} \bullet)
    =
      2
  \\[1ex]
      3 \leftsub 5
    =
      (\defdes k {3 = 5 + k} \bullet)
    =
      \bullet
  \\[1ex]
      3 \leftsub \emptyset
    =
      (\defdes k {3 = \emptyset + k} \bullet)
    =
      \bullet
  \end{array}
$$%

\medskip
\noindent\textbf{Function Application and Some Other Function Operators.}
\newcommand{\ap}{\mathbin{`}}%
Define the \emph{function application} infix operator
$(\ap)::\domain 0\Fun\domain 0\Fun\domain 0$ such that if $f:\Func$ and $x\in\dom f$, then $f\ap x$
is the value of $f$ at $x$, and otherwise $f\ap x=\bullet$:
$$%
  (\ap) = (\lam {f\,x} {\defdes y {\fapp f x y} {\bullet}})
$$%
Define an operator $(\dot\lambda)$ that when given a domain $x:\Set$ extracts from an operator
$t::\domain 0\Fun\domain 0$ a function of type $\domain 0$:
$$%
    \dot\lambda
  =
    (\lam {x\, t}{\mkfun x {(\lam {b\,c} {b \mathbin{\in} x \land c \mathbin{=} \app t b})}})
$$%
We furthermore adopt the following nice notation:
\(
    (\lamfun{b \in x}{B})
  \abbreviates
    (\apptwo{\dot\lambda}{x}{(\lam b B)})
\).
Define an operator to lift a binary operator on objects of type $\domain 0$
to a binary operator on functions:
$$%
    \mathsf{lift}
  = \lam{o\, f\, g}
        {(\binddot{\dot\lambda}
                  {b \mathbin{\in} {\dom f} \cap {\dom g}}
                  {\apptwo o{(f \ap b)}{(g \ap b)}})}
$$%
Define an operator
$\mathsf{FunRet}::(\domain 0\Fun\bool)\Fun\domain 0\Fun\bool$
that takes a soft type $P$ and yields the soft type of functions whose return values always satisfy
$P$:
$$%
    \mathsf{FunRet}
  = (\lam{p\,f}{f : \Func \wedge \ran f : \SetOf{p}})
$$%

\medskip
\noindent\textbf{Pointwise Left-subtraction on Ordinal-valued Functions.}
A function $f:\Func$ is \emph{ordinal-valued} iff $f : \app{\mathsf{FunRet}}{\Ord}$.
Thus, $\appfour{\mathsf{lift}}{(\leftsub)}fg$ computes the pointwise left-subtraction of
ordinal-valued functions $f,g: \app{\mathsf{FunRet}}{\Ord}$.
For example, consider functions $f,g$
  with $\dom f$ and $\dom g$ being $\predSet\omega$ (the set of all ordinals less than $\omega$),
  where $f \ap j = j + 2$ and $g \ap j = j + 1$ for all $j \in \predSet\omega$.
Clearly $f,g$ are ordinal-valued.
Hence for any ordinal $j < \omega$,
$$%
  \begin{array}{r@{\:=\:}l}
         (\appfour{\mathsf{lift}}{(\leftsub)}fg) \ap j
      &  (\lamfun{b\in \dom f \cap \dom g}{(f \ap b) \leftsub (g \ap b)}) \ap j
    \\
      &  (f \ap j \leftsub g \ap j) 
    = 
      (j + 2) \leftsub (j + 1)
    \\
      &  2 \leftsub 1
    = 
      1
  \end{array}
$$

\medskip
\noindent\textbf{Combining Operators.}
Define an operator
\(
     {\mathsf{override}}
\)
that takes two binary operators $o_1,o_2:: \domain 0\Fun\domain 0\Fun\domain 0$ and two predicates
$t_1,t_2::\domain 0\Fun\domain 0\Fun\bool$ saying when to use each operator and builds a new
operator by combining them:
$$%
    \appfour{\mathsf{override}}{t_1}{o_1}{t_2}{o_2}
  =
    (\lambda\:x\,y.\,
                    \begin{array}[t]{@{}l@{}}
                       \mathsf{if}\ (\apptwo{t_1}xy)\ \mathsf{then}\ (\apptwo{o_1}xy)
                    \\ \mathsf{else}\ \mathsf{if}\ (\apptwo{t_2}xy)\ \mathsf{then}\ (\apptwo{o_2}xy)
                    \\ \mathsf{else}\ \bullet)
                    \end{array}
$$%
Define an infix operator $(-)$ that combines ordinal left-subtraction
with ordinal left-subtraction lifted to ordinal-valued functions:
$$%
    ({-})
  = \mathsf{override}
          \,{(\lam{x\,y}{\ x,y:\Ord})}
          \,{(\leftsub)}
          \,{(\lam{x\,y}{\ x,y:\app{\mathsf{FunRet}}{\Ord}})}
          \,{(\app{\mathsf{lift}}{(\leftsub)})}
$$%
If $i,j$ are ordinals, then $i - j=i \leftsub j$.
If $f,g$ are ordinal-valued functions,
  then $f-g$ is the function such that
  if $j \in \dom f$ and $j \in \dom g$,
  then $(f - g) \ap j = (f \ap j) \leftsub (g \ap j)$, and otherwise
  $(f - g) \ap j = \bullet$.
Because $\Ord\inter\Func=\bot$ is an axiom of $\ZFplus$, we know that at most one of
$x,y\mathbin:\Ord$ and
$x,y\mathbin:{\app{\mathsf{FunRet}}{\Ord}}$ can
be true, so there is no possibility of $(-)$ using $(\leftsub)$ where the user might have intended
instead for $(-)$ to use $(\app{\mathsf{lift}}{(\leftsub)})$.

\section{Building a Model for ZF$^+$ in ZFC}%
  \label{sec:zfc-zfplus}

Normally, using our example GST $\ZFplus$ is done in a type that it is instantiated at.
Such a type could be gained by including the axioms of $\ZFplus$ in the hypotheses of deductions
  (i.e., $\ZFplus_{\msf{axioms}}\typSub{\domain 0} \subseteq \Gamma$).
For confidence in $\ZFplus$'s consistency, we opt to \emph{define} $\domain 0$ in terms of a model
built in $\mathbf{V}$, a type axiomatized by Paulson's $\ZFCinHOL$.%
\footnote{Interested readers can see the definition of the $\ZFplus$ class in
  \texttt{GST\_Features.thy}, and the type definition of $\domain 0$ and instantiation of the
  $\ZFplus$ class in \texttt{Founder/Test.thy}.}
Models of GSTs are cumulative hierarchies of sets defined via ordinal recursion.
We build models of GSTs within other GSTs using classes called \emph{model components} that
correspond to features.
We use Isabelle/ML code to generate terms, theorems, and proof states that assist
   in instantiating these classes and using them as interpretations of GST features.
Details of our approach may be seen in the appendix of the longer version of this paper.

The only axioms (i.e., excluding simple definitions) used for this are 
  those of $\HOL$ and $\ZFCinHOL$.
From this point, only \emph{definitional mechanisms}
   (as defined by Kun\v{c}ar and Popescu~\cite{kuncar2017})
   are used to define $\ZFplus$ in $\domain 0$
  (i.e., class instantiation, \texttt{typedef}, \texttt{lift\_definition}).
Hence, if $\HOL$ and $\ZFCinHOL$ are consistent, then so are the axioms of $\ZFplus$.

\subsection{Building a model in $V$}
\label{sub:building_a_model_in_v}

To build a model in $\mathbf{V}$, our method needs the $\GZF$, $\Ordinal$, $\cOrdRec$, and
$\Function$ classes to be instantiated at $\mathbf{V}$.

$\ZFCinHOL$ provides axioms for set theory,
  and supports von Neumann ordinals and transfinite recursion.
Hence instantiations for $\GZF$, $\Ordinal$, $\cOrdRec$
are achieved by choosing correct definitions for each parameter,
and then using automatic proof methods to discharge proof obligations.
For $\Function$, we have a generic result that this
class may be instantiated in any GST with the $\GZF$ feature,
using sets of Kuratowski ordered pairs.

We define a model in $\mathbf{V}$ and a soft type $\M :: \mathbf{V} \Fun \bool$
of all objects in the model.
We then do an Isabelle/HOL type definition to define $\domain 0$ as a type isomorphic to the model
of $\ZFplus$ built in $\mathbf{V}$
(i.e., $\app{\texttt{typedef}}{\domain 0} = \app{\mathsf{Collect}}{\M}$).%
\footnote{%
  The operator $\mathsf{Collect}$ converts a predicate to a HOL-set as required by
  \texttt{typedef}.}

\subsection{Instantiating ZF$^+$ in $\domain 0$}
\label{sub:instantiating_zf_in_domain_0}

To instantiate the $\ZFplus$ class at $\domain 0$,
  we must provide instantiations for each parameter of $\ZFplus$,
  and prove each axiom.
This is achieved with help from the
Lifting and Transfer~\cite{kuncar2013} packages.
The parameters of each feature of $\ZFplus$ ---
namely $\GZF$, $\Ordinal$, $\Function$ and $\Exc$ ---
are instantiated using Isabelle/ML code calling the Lifting package
to lift constants acting on the model in $\mathbf{V}$ to $\domain 0$.
To prove the axioms of each feature,
   the Transfer package is used to transform each subgoal
   into an equivalent statement about the model.
Our approach only needs the transferred versions of a feature's axioms to be proved once --- the
proofs may be reused when building a model of any GST using that feature.
The axioms of $\ZFplus$ generated in the feature combination process
   (e.g., $\cover$, $\disjoint$) must also be proved on $\domain 0$.
This is also achieved with Transfer,
   but these statements are proved manually.

\section{Related and Future Work}

\noindent\textbf{Other Related Work.}
Two bodies of work are notable in being similar in spirit to the way we assemble GSTs, specifically
work by Farmer et al.\ on ``little theories''~\cite{Far+Gut+Tha:CADE-1992}, which
emphasizes small theories that are then connected together by morphisms, and work led by Rabe on the
MMT framework~\cite{Rab:LU:2015}, which emphasizes combining even smaller theory bits with morphisms.

\medskip
\noindent\textbf{Future Work.}
We aim to further develop generalized set theories and our Isabelle/HOL formalization in the
following directions.
\textit{Importing features from Isabelle/HOL.}
The ``ZFC in HOL'' development that we bootstrap from provides support (packaged up in two type
classes) for obtaining structures within the set theory with the same behavior as other Isabelle/HOL
types.
We aim to be able to pass such structures on to GSTs we build in the form of extra features.
\noindent\textit{Foundation.}
We have not yet implemented the axiom of foundation, which states that there are no infinite chains
going by steps from an object to any of its ``children''.
Many use cases do not need this, so it is not crucial, but some of our future work
will depend on it.
We plan to handle this as described in previous work
  on formalizing GSTs~\cite{Dun+Wel+Kam:CICM-2021}.
\noindent\textit{Overloading.}
Isabelle supports overloading (via classes or \emph{ad hoc} commands) so symbols can have different
meanings at different types.
However, we want to give symbols different meanings for different features within one type, a GST's
domain.
For example, we want machinery that will support overloading $\in$ to work both with $\Set$ objects
and groups within a GST.
We also want to allow symbols for set operations like $\in$ to be used both with soft types (which
are close to Isabelle/HOL sets) and $\Set$ objects in a GST.
\noindent\textit{Universes.}
To support some applications in category theory and also for building models of complicated type
theories to embed them within a GST, we aim to add a feature that adds some kind of universe axiom
like in TG, so those who need it could have a much larger and more powerful GST.
\noindent\textit{Automation.}
Others are working on automation of finding and using soft types and we aim to build on any
progress they make.
Our model-building process currently needs lots of manual proving, but much can be automated and/or
given a nicer interface.

\iffalse
  \printbibliography
\else
  \bibliographystyle{jbwc}
  \bibliography{dunne-phd}
\fi

\appendix

\section{Appendix: Model Building Kit}

We believe the material in this appendix is secondary and should not be needed
for understanding the main points of the paper.
This appendix is primarily aimed at people who might like to customize the features we provide and
therefore
need to figure out how to generate a model for their customized GST to have confidence in its
consistency.

\subsection{The Base Model}

\begin{boxfigure}[t!]{fig:tagging}
  {Definition of $\bfOPair$, $\Tagging$, $\ModelBase$, and introduction rules for $\cTier$.}
  $$
   \begin{array}{r@{\,}l}
    \Pair_\consts \abbreviates
    [ &
      \defaultval\Pair :: \alpha,\:
      \Pair :: \alpha \Fun \bool,\:
      \cpair :: \alpha \Fun \alpha \Fun \alpha
    \,]
  \\[.5ex]
    \multicolumn{2}{@{}l@{}}{
      \mbox{Henceforth, the notation $\prettypair BC$ stands for $\pair BC$.}
    }
  \\[.5ex]
    \Pair_\axioms \abbreviates
  [ &
    \cpair : \PairMem \fun \PairMem \fun \Pair,
  \\ &
    \all {b,c,u,v : \PairMem}
      {\prettypair bc = \prettypair uv \iff (b = u \land c = v)},
  \\ &
    \all {p : \Pair}
      {\ex {b,c} {p = \prettypair bc}}
  \:]
  \\
    \Pair_\defs \abbreviates
    [ &
      \PairMem = (\lam b {\ex {p : \Pair} {\ex c {p = \prettypair bc \lor p = \prettypair cb}}})
    \:]
  \\
    \OPair \abbreviates &
      \mlist{\nil, \Pair_\consts, \Pair_\axioms, \Pair_\defs}
  \\
    \bfOPair \abbreviates &
      \mlist{\OPair, \Pair, \PairMem, \defaultval\Pair}
  \end{array}
  $$
  \hrule
  $$
    \predisplaysize=0pt\relax
   \setof{ b \in X \mid \varphi}
      \abbreviates \Repl X {(\lam {b\,c} {b = c \land \varphi})},
   \quad
   \setof{B \mid b \in X}
   \abbreviates \Repl X {(\lam {b\,c} {c = B})},
  $$
  $$
   \setof{B \mid u < C} \abbreviates
      \setof{B \mid u \in \predSet C},
  $$
  $$
   (\lam {u < B} {C}) \abbreviates \mkfun {(\predSet B)} {(\lam {u\,v} {v = C})}
  $$
  $$\begin{array}{c}
     \begin{array}{rll}
    \Tagging_\axioms & \abbreviates [
    &
      \SetMem \subtyp \PairMem, \Pair \subtyp \SetMem \:]
    \\
    \Tagging_\defs & \abbreviates [
    &
      \Tag = \Ord \inter (\lam i {i < \omega}),
    \\ & &
      \tagmap = (\lam {i\,x}{\setof{\prettypair ib \mid b \in x}}),
    \\ & &
      \cdisj = (\lam {f} {\Union {\setof {i \tagmap \app f i \mid i < \omega}}})
    \:]
  \end{array}
  \\ 
   \Tagging \abbreviates 
      \mlist{\mlist{\OPair, \cOrdRec}, \nil, \Tagging_\axioms, \Tagging_\defs}
   \end{array}$$
\hrule
     $$\begin{array}{rll}
      \ModelBase_\consts & \abbreviates [
    &
      \cTierImp :: \alpha \Fun \alpha \Fun \alpha \Fun \alpha,\:
      \Excluded :: \alpha \Fun \alpha \Fun \bool
    \:]
  \\
      \ModelBase_\axioms & \abbreviates [
    &
      \cTierImp : (\Tag \fun {\Ord \fun \Set \fun \Set}),
    \\ & &
      \cTierImp : (\Tag \fun {\Limit \fun \Function \fun \Set})
    \:]
    \\
    \ModelBase_\defs & \abbreviates [
    &
      \upair = (\lam {b\, c :: \alpha} {\ldots}),
    \\ & &
      {\cup} = (\lam {b\,c} {\Union {\apptwo\upair b c}}),
    \\ & &
      {\ominus} = (\lam {x\, p} {\setof{b \in x \mid \lnot b : p}}),
    \\ & &
      \cTier_\zero = \disj (\lam i {\TierImp i 0 \Emp}),
    \\ & &
      \cTier_\csucc = (\lam {j\,x} {x \cup \disj (\lam i {\TierImp i j {(x \ominus \Ex_i)})}}),
    \\ & &
      \cTier_\limit =
        (\lam {u\, f}
          {(\Union \setof {\app f j \mid j < u})
    \\ & & \qquad\qquad\quad
              \cup\, (\disj (\lam i {\TierImp i u {(\lam {j<u} {\app f j \ominus \Ex_i})}}) })),
    \\ & &
      \cTier = \OrdRec{\cTier_\limit}{\cTier_\csucc}{\cTier_\zero},
    \\ & &
      \M = (\lam b {\ex {i : \Ord} {b \in \Tier i}})
    \:]
  \end{array}
$$
$$
  \begin{array}{l}
    \ModelBase \abbreviates
      \mlist{ \mlist{\Tagging, \Function},
      \ModelBase_\consts, \ModelBase_\axioms, \ModelBase_\defs}
  \end{array}
$$
\hrule
$$\predisplaysize=0pt\relax
  \begin{array}{r@{\:\:}l}
  \mathlabel{zero}
&
  \all {i : \Tag} {\all b
    {b \in \TierImp i 0 \Emp  \imp \prettypair ib \in \Tier 0}}
\\[1mm]
  \mathlabel{succ1}
&
  \all {j : \Ord} {\all b {b \in \Tier j \imp b \in \Tier (\succ j)}}
\\
  \mathlabel{succ2}
&
  \all {i : \Tag} {\all {j : \Ord} {\all b
    {b \in \TierImp{i}{(\succ j)}{(\Tier j \ominus \Ex_i)}}
      \imp b \in \Tier j}}
\\[1mm]
  \mathlabel{lim1}
&
  \all{u : \Limit}{\all {j : \Ord}
    {j < u \imp {(\all b {b \in \Tier j \imp b \in \Tier u})}}}
\\
  \mathlabel{lim2}
&
  \all {i : \Tag} {\all {u : \Limit} {\all b
    {b \in \TierImp i u {(\lam {j<u} {\Tier j \ominus \Ex_i})}
      \imp b \in \Tier u}}}
\end{array}$$
\end{boxfigure}

As with most set theories,
  models of GSTs are built as cumulative hierarchies of sets indexed by ordinals.
Unlike pure set theories, we cater to the possibility
  that some objects in the domain are not sets.
We implement this by requiring that models consist entirely of \emph{tagged objects}
   $\prettypair JB$
  where $J$ is a finite ordinal \emph{tag}, with \emph{data} $B$.

To support building and reasoning on tagged objects,
   \autoref{fig:tagging} defines the $\bfOPair$ feature and the $\Tagging$ class.
$\bfOPair$ provides ordered pairs.
The formulas in $\Tagging_\axioms$ assume that the structure on $\SetMem$, $\PairMem$, and $\Pair$
  allows pairs to be formed from set members and allows pairs to be members of sets.
$\Tagging_\defs$ provides the soft type $\Tag$ of finite ordinals,
  tag-map $\tagmap$, and disjoint union $\cdisj$.
The development $\Tagging$ can be seen in \texttt{Model/Tagging.thy}.

Unlike typical models of set theories with non-structured atoms (ZFU, KPU), we deal with objects
that can not be added at the first tier in the hierarchy because they depend on objects in the
hierarchy, and also objects where it might not make sense to add them at the first tier because
there are more than a proper class of them (in our case we can not handle this because we represent
each tier in the hierarchy by a set).
For example, $\ZFplus$ has the $\bfGZF$ and $\bfOPair$ features.
  When building a model, the tagged objects representing ordered pairs
  need to be able to include sets formed at any tier, so at any tier there must be more ordered
  pairs that are not in the tier and still need to be added.
Instead, tagged objects for pairs must be added at each tier along with those for sets.
The same problem is faced with other structured
  proper-class-sized features like $\bfFunction$, and $\bfOrdinal$.
Thus, we must provide specifications for which objects should be added to each tier.
Another important requirement is the possibility for certain objects to be
  excluded from the internal structure of a features objects.
Both requirements are satisfied by our model operator $\cTier$,
  whose definition makes use of
   a table of operations $\cTierImp$ (where ``Imp'' is for ``Implementation''), and
   a tag-indexed family of soft types $\Excluded$.
Both $\cTierImp$ and $\Excluded$ are specified by the user
   when building a model in a concrete type, using terms
   generated by Isabelle/ML code.
For each tag $i$ and ordinal $j$, $\apptwo\cTierImp i j$
   specifies an operation used for adding a set of objects to $\Tier j$,
and $\app\Excluded i$ specifies the objects ignored when building new tiers.

\Autoref{fig:tagging} defines $\ModelBase$,
   where $\Ex_i \abbreviates {\app \Excluded i}$.
Soft type assumptions are made for $\cTierImp$, 
  and the definition of $\cTier$ and $\M$ are given.
The development of $\ModelBase$ is in
  \texttt{ModelKit/ModelBase.thy}, \texttt{ModelKit/ModelTypes.thy}.

\begin{boxfigure}[t!]{fig:mcomp-class}
  {Building classes for model components, and an example model component.}
  $$
    \text{Given $\mcompvar \meq [\kappa, \mathcal{D},\constructor, \Upphi, \Uppsi\:],
  \quad
    \constructor \meq [Z,S,L]$, then}
  $$
  $$
  \begin{array}{l@{\,}c@{\,}l}

    \mcomp_\consts(\mcompvar)
  &
    \abbreviates [
  &
    \kappa :: \alpha, \defaultval\kappa :: \alpha
  \:]
  \\
    \mcomp_\axioms(\mcompvar)
  &
    \abbreviates [
  &
    \kappa : \Tag,
  \quad
    \defaultval\kappa : \M,
  \quad
    \all x {\TierImp \kappa 0 x = Z},
  \\ & &
    \all {j : \Ord} {\all x
      {\TierImp \kappa {(\succ j)} x = \apptwo{S}{(\succ j)}{x}}}
  \\ & &
    \all {u : \Limit} {\all f
      {\TierImp \kappa u f = \apptwo{L}{u}{f}}}
  \:] \append \Upphi
  \end{array}
  $$
  $$
    \mcomp(\mcompvar) \abbreviates
      \mlist{\mlist{ \ttag(\mcompvar') \mid \mcompvar' \listin \mathcal{D}},
        \mcomp_\consts, \mcomp_\axioms, \Uppsi}
  $$
\hrule
  $$
    \predisplaysize=0pt\relax
  \begin{array}{r@{\,}c@{\,}l}
    \mGZF_\kons
  &
     \abbreviates [
  &
    \Emp,\:(\lam {j\,x} {\Pow x}),\:(\lam {u\,x} {\Emp})]
  \\
    \mGZF_\excluded
  &
    \abbreviates [
  &
    \all x {\lnot {\prettypair\set x : \app\Excluded\set}}
  \:]
  \\
    \mGZF_\defs
  &
    \abbreviates [
  &
    \mSet = \M \inter (\lam {b} {\fst b = \set}),
  \\  & &
    {\mmem} = (\lam {b\,y} {y : \mSet \land b \in \snd y}),
  \\ & &
    \mSetMem = (\lam b {\ex {x : {\mSet}} {b \mmem x}}),
  \\ & &
    \mSetOf = (\lam {p\,x} {x : \mSet \land (\all {b} {b \mmem x \imp b : p})})
  \\ & &
    \cmUnion' = (\lam x {\prettypair\set{\Union \setof{ \snd y \mid y \in \snd x}}})
  \\  & &
    \cmUnion =
      (\lamelse {x : \mSetOf \mSet}{\app{\cmUnion'}{x}}{\defaultval\set}),
  \\ & &
    \cmPow' =
      (\lam {x} {\prettypair\set{\set \tagmap {\Pow {(\snd x)}}}}),
  \\ & &
    \cmPow =
      (\lamelse {x : \mSet} {\app{\cmPow'}{x}} {\defaultval\set}),
  \\ & &
    \mEmp = \prettypair\set\emptyset,
  \\ & &
    \cmSucc =
      (\lamelse {x : \mSet}
         {\prettypair \set{\lbrace x,
            \,\prettypair{\set}{\{x\}}\rbrace}} {\defaultval\set}),
  \\ & &
    \mSetOrd = \OrdRec
      {(\lam {u\,f} {\prettypair\set{\setof{\app f j \mid j < u}}})}
      {(\lam {j\, x} {\mSucc x})}
      {\mEmp}
  \\ & &
    \mInf = \app \mSetOrd \omega
  \\ & &
    \cmReplPred =
      (\lam {x\,p} {\all b {b \mmem x \imp {\atMostOne {b : \mSetMem} {\apptwo p a b}}}}),
  \\ & &
    \cmRepl' =
      (\lam {x\,p}
        {\prettypair\set{\Repl {(\snd x)} {(\lam {b\,c} {\apptwo p a b \land b : \mSetMem})}}}),
  \\ & &
    \mRepl =
      (\lamelse {x : \mSet, p : \mReplPred x}
        {\apptwo{\cmRepl'}{x}{p}} {\defaultval\set})
 \:]
  \end{array}
$$
$$
  \mGZF \abbreviates
    \mlist{\set, \nil, \mGZF_\kons, \mGZF_\excluded, \mGZF_\defs}
$$
\end{boxfigure}

\subsection{Model Components}

A \emph{constructor} is a triple $\constructor \meq [Z, S, L]$, where
  $Z : \Set$,
  $S : \Ord \fun \Set \fun \Set$,
  $L : \Limit \fun \Function \fun \Set$.%
\footnote{$\constructor$ is Fraktur letter C.}
A \emph{model component} is a tuple
  $\mcompvar \meq [\kappa, \mathcal{D},\constructor, \Upphi]$
such that:
  $\kappa :: \alpha$ is the \emph{tag} of $\mcompvar$
    (referred to by $\ttag(\mcompvar)$),
  $\mathcal{D}$ is a list of model components called the \emph{dependencies},
  $\constructor$ is a constructor,
  $\Upphi$ is a list of formulas used for constraining $\Excluded$, and
  $\Uppsi$ is a list of simple definitions.
\Autoref{fig:mcomp-class} defines the meta-level function $\mcomp$
  for creating classes from model components,
  and provides an example model component for the $\bfGZF$ feature.

\subsection{Parameter Morphisms}

A \emph{parameter morphism} is a function $\eta : \Const \to \Const$
  mapping constants given by a feature to constants given by a model component.
Parameter morphisms are used for automating our uses of the Lifting \& Transfer (LT)~\cite{kuncar2013} package.
  Namely, we use them to prove generic facts which correspond to the goals
  generated by LT when lifting a model into a new type.
An example of a parameter morphism is $\mGZF_\map$, where:
$$
\begin{array}{r@{\,}c@{\,}l}
  \mGZF_\map 
  & 
    \abbreviates [ 
  &
    \Set \mapsto \mSet,\:
    {\in} \mapsto {\mmem},\:
    \cUnion \mapsto \cmUnion,\:
    \cPow \mapsto \cmPow,\:
    \Emp \mapsto \mEmp,\:
  \\ & & 
    \cSucc \mapsto \cmSucc,\:
    \Inf \mapsto \mInf,\:
    \cRepl \mapsto \cmRepl,
  \\ & &
    \SetMem \mapsto \mSetMem,\:
    \cSetOf \mapsto \cmSetOf,\:
    \cReplPred \mapsto \cmReplPred
  \:]
\end{array}
$$
We provide Isabelle/Isar commands
 $\translate$ and $\resp$.
The command $\translate(\feat, \eta)$
  sets up a proof state with the translation of each axiom of
  a feature $\feat$ with respect to the mapping $\eta$.
Our translation algorithm works by descending into the structure of formulas,
  replacing constants $\kappa$ in the domain of $\eta$ by $\eta(\kappa)$.
Quantifiers are replaced by model-bounded versions
  (e.g., $\ex {x : P} {\varphi}$ becomes ${\ex {x : \M \inter P} {\varphi}}$),
  and all other terms are untouched.
For example, when performing $\translate(\bfGZF, \mGZF_\map)$,
  the empty set axiom $\all {b} {\lnot (b \in \Emp)}$ becomes
  $\all {b : \M} {\lnot (b \mmem \mEmp)}$.

The command $\resp(\feat, \eta)$ generates goals for \emph{respectfulness theorems}
  for each constant in the image of $\eta$ on $\params(\feat)$.
Respectfulness theorems are proof obligations required whenever LT's
  \texttt{lift\_definition} command is invoked.
As an example, $\resp(\bfGZF, \mGZF_\map)$,
  requires proving $\all x {x : \M \imp \mPow x : \M}$.

Implementations
  can be found in \texttt{ModelKit/Tools/translate\_axioms.ML}.
Uses of both commands can be seen in \texttt{GZF/Model/GZF\_Model.thy},
  and the resulting theorems are used for building a model
  for $\ZFplus$ in \texttt{Founder/Test.thy}.

\end{document}